\newcommand\orc[1]{\href{https://orcid.org/#1}{\includegraphics[width=3mm]{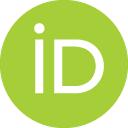}}}

\documentclass{aa}  

\usepackage{natbib}
\usepackage{verbatim}
\usepackage{bm}
\usepackage{graphicx}
\usepackage{txfonts}
\usepackage[colorlinks=true,
            linkcolor=blue,
            citecolor=blue,
            urlcolor=blue]{hyperref}
\begin{document}

   \title{Asteroseismic imprints of mass transfer in binary stars: probing the interiors of donors and accretors with gravity and acoustic modes}

   \subtitle{ }
  
   \author{Tao Wu\inst{1,2,3,4,5} \orc{0000-0001-6832-4325}
          \and
     Zhao Guo \inst{6,7} \orc{0000-0003-1822-7126}
     \and
     Yan Li\inst{1,2,3,4,5} \orc{0000-0002-1424-3164}
     }
   \institute{Yunnan Observatories, Chinese Academy of Sciences, 396 Yangfangwang, Guandu District, Kunming, 650216, P. R. China\\
              \email{wutao@ynao.ac.cn;ly@ynao.ac.cn}
                          \and
             Key Laboratory for the Structure and Evolution of Celestial Objects, Chinese Academy of Sciences, 396 Yangfangwang, Guandu District, Kunming, 650216, P. R. China
\and
International Centre of Supernovae, Yunnan Key Laboratory, 396 Yangfangwang, Guandu District, Kunming, 650216, P. R. China
\and 
University of Chinese Academy of Sciences, Beijing, 100049, P. R. China
\and
Center for Astronomical Mega-Science, Chinese Academy of Sciences, 20A Datun Road, Chaoyang District, Beijing, 100012, P. R. China
\and
Institute of Astronomy (IvS), KU Leuven, Celestijnenlaan 200D, B-3001 Leuven, Belgium
\and
Department of Applied Mathematics, School of Mathematics, University of Leeds, Leeds LS2 9JT, UK
    \\          
    \email{z.guo2@leeds.ac.uk}
}

   \date{Received Nov 5, 2025; accepted Nov 17, 2025}

% \abstract{}{}{}{}{} 
% 5 {} token are mandatory
 
  \abstract
   {The synergy between close binary stars and asteroseismology enables constraints on mass-transfer episodes and their consequences for internal structure, rotation profiles, and oscillation modes. }
   {We investigate how mass accretion and donation in close binaries affects the internal structure and oscillation modes of main-sequence stars.}
   {Building on the established relation between the Brunt–Väisälä (buoyancy) glitch and the Fourier spectra of g-mode period spacings, we quantitatively explain the origins of the g-mode period-spacing differences between single-star and mass-accretion/donation models of intermediate-mass stars ($M = 2.0$, $3.0$, and $4.5~{\rm M_{\odot}}$). In particular, the hydrogen mass fraction profiles $X$ of the donor model show two chemical gradient regions, which results in a double-peaked Brunt–Väisälä profile. The presence of additional buoyancy glitches gives rise to further periodic modulations in the g-mode period spacings.}
   {Mass-accretion–induced changes in the chemical profile create sharp features in the buoyancy frequency, which modify both the amplitudes and frequencies of the g-mode period-spacing variations.
This behavior resembles that produced by multiple chemical transition zones in compact pulsators such as white dwarfs and sub-dwarf B stars.
Similarly, for acoustic modes in the $M=1~{\rm M_{\odot}}$ solar-like models, we attribute the differences in frequency-separation ratios between single-star and mass-donor models to the variations in the internal sound-speed gradient (acoustic glitches). We discuss future prospects for using asteroseismology to discover the mass-transfer products and constrain the mass-transfer processes in binary star evolution. } 
   {}


  \maketitle

\section{Introduction}
Binary interaction is a cornerstone of stellar astrophysics, as a substantial fraction of stars form and evolve in binary or multiple systems \citep{San12,Moe17}. For such systems, the exchange of mass and angular momentum profoundly alters the internal evolution of both components and the orbital configuration \citep{Pac71,Lan12,DeM13}.

The dominant interaction process is mass transfer, which can occur through several distinct channels. The most common is Roche lobe overflow (RLOF), where a star expands to fill its Roche equipotential surface and material flows through the inner Lagrangian point to the companion \citep{Bon44,Kol90}. RLOF episodes are traditionally classified as Case A, B, or C depending on whether they begin during core hydrogen burning, shell hydrogen burning, or core helium burning, respectively \citep{Kip67, Lau70, Pol94}. Another important channel is stellar-wind mass transfer \citep{Smi14}

Mass transfer strongly influences the structure and evolution of both stars. For the mass donor, rapid mass loss can drive the star out of thermal equilibrium, expose nuclear-processed layers, and lead to the formation of stripped helium stars \citep{Lap20}, hot sub-dwarfs \citep{Han01}, or white dwarfs \citep{Ibe93, Wil04, Cor19}. 
For the mass gainer, accretion increases its mass, luminosity, and angular momentum, often inducing differential rotation and chemical mixing \citep{Neo77, Pack81, Lan12}. The accretor may rejuvenate its structure via convective core growth and display an apparent age younger than its true evolutionary state \citep{Van98}. These effects have profound consequences, contributing to the formation of blue stragglers in clusters \citep{Che09}, cataclysmic variables and X-ray binaries \citep{Rap83, Tau06, Gie13}, the emergence of rapidly rotating Be stars \citep{Riv13}, compact binary mergers that serve as progenitors of Type Ia supernovae and gravitational-wave sources \citep{Web84, Han95, Bel02}, and the characteristics of runaway stars produced by supernova ejections \citep{Per12}.

Recent advances in asteroseismology \citep{Aer10,Bas17} have opened a powerful new window into diagnosing the internal consequences of binary mass transfer. The oscillation spectra of intermediate- and high-mass stars provide direct probes of their internal chemical and thermal structure, which retain the imprints of past accretion or mass-loss episodes. In particular, gravity-mode (g-mode) pulsations are highly sensitive to the sharp variations in the Brunt–Väisälä (hereafter Brunt) frequency ($N^2=g\left (\frac{d\ln p}{dr}\frac{1}{\Gamma_1}-\frac{d\ln\rho}{dr}\right )$) that arise from composition gradients left behind by nuclear burning or incomplete mixing \citep{Mig08}. In post-mass-transfer stars, the altered hydrogen abundance profiles and non-standard core-envelope structures produce distinctive buoyancy glitches and period-spacing modulations in the g-mode spectra \citep{Wu18, Guo21, Zha23, Hat23, Farr2024, Guo25}. These features encode information about the extent of the convective core, the efficiency of mixing at the chemical gradient zone, and the degree of structural rejuvenation in the mass gainer. Consequently, seismic analysis provides a unique means to constrain the internal mixing and evolutionary history of stars affected by binary interaction—offering direct observational tests of classical theoretical predictions for accretors and donors.

We are beginning to uncover how the synergy between mass transfer and asteroseismology can reveal the internal fingerprints of binary interaction. We briefly summarize the observationally identified post-mass-transfer stars with oscillations.

{Post-mass-transfer main-sequence pulsators:}

\citet{Guo17} found a post-mass transfer high-frequency $\delta$ Scut pulsator with a low-mass helium white dwarf companion the eclipsing binary KIC 8262223, {and the main-sequence + He white dwarf combination can be explained by the standard formation process of EL CVn stars} \citep{Che17}. They suggest that the high-frequency p modes are likely the effect of rejuvenation due to mass accretion. A similar system, TT Hor was studied by \citet{Str18}. \citet{Mis21} and \citet{Mis22} studied similar post-mass transfer binaries with p-mode pulsations in KIC 10661783 and AB Cas. They performed a binary-star-evolution modeling and examined the p mode pulsation frequencies and mode excitation. \citet{Guo17b} explained the observed g-mode period spacing and p-mode frequencies in a post-mass-transfer A-type star in an eclipsing binary KIC 9592855. Oscillating Algol (oEA) binaries have been studied by \citet{Mkr04, Mkr18} and many show oscillations as well as accretion-driven variabilities.

{Post-mass-transfer red giants:}
\citet{Li22}, \citet{Deh22}, and \cite{LiG2024AA} identified a group of red giant stars which deviate from the branch of core-He-burning {red clump stars} on the period spacing-frequency diagram. They found they are likely
post-mass-transfer products, which can explain their observed location.

{White dwarf pulsators with mass transfer:}
\citet{Arr06} shows that accreting white dwarfs in cataclysmic variables (CVs)
can develop g-mode pulsation instabilities over a range of effective temperatures.
\citet{Kum23} studied the effects of temperature change and rotation on the g-mode behavior due to accretion in white dwarf models. Extremely Low-Mass (ELM) white dwarf variables ($M \leq 0.2~{\rm M_{\odot}}$) are products of binary evolution, where mass transfer strips the envelope before helium ignition on the red giant branch. They show non-radial p and g modes \citep{MAX13,gIA16,Ist16,Ist16a}.

{Post-mass-transfer RR Lyrae and Cepheid pulsators:}
Post-mass-transfer RR Lyrae and Cepheid pulsators have been found by \citet{Pie12} and \citet{Pil17}, respectively. \citet{Gau17} and \citet{Kar17} studied the binary evolution channel to form anomalous Cepheids via RLOF and merger-like evolution. 

A review on the mass-transferring and post-mass-transfer binary with pulsations in B, A, F-type stars and others can be found in \citet{Guo21} and \citet{Sou25}.

{On the theoretical/modeling side of mass transfer studies}, \citet{Neo77} modeled how steady mass accretion alters the internal evolution of a main-sequence star, showing that accretion accelerates helium-core growth and produces structural differences from single-star evolution of the same mass.
\citet{Wag24} show that mass accretion onto a Slowly Pulsating B star (SPB, $3.5~{\rm M_{\odot}}$) leaves persistent g-mode signatures—stronger oscillatory period-spacing patterns from altered chemical gradients during rejuvenation. \citet{Mis25b} demonstrate that accretion in a $10~{\rm M_{\odot}}$ “$\beta$ Cep” binary strongly reshapes the stellar interior by expanding the convective core, creating local density bumps, and modifying chemical gradients, producing distinct asteroseismic features in both g-mode spacings and p-mode separations. Their weight-function analysis pinpoints the spatial origin of these frequency differences. \citet{Ren21} and \citet{Mis25a} find that off-center convective zones naturally develop in accreting stars.
\citet{Ren21} also showed that the surface of the accretor is polluted by CNO-processed material donated by the companion.

{Similar to mass transfer, stellar mergers can also produce stellar remnants that exhibit oscillations.} 
\citet{Rui21} studied the fingerprints of stellar merger in low-mass red giant stars. 
For massive stars, \citet{Hen24} showed that merger products ($6 +2.4~{\rm M_{\odot}}$ and $9+6.3~{\rm M_{\odot}}$) differ seismically from single stars of similar HRD position, with lower $\Pi_0$ (asymptotic period spacing) values, 
double g-mode cavities, and deep period spacing dips - making them distinguishable with asteroseismology. Based on a 3D MHD simulations of a stellar merger and relaxing into a 1D model in MESA, \citet{Hen25} show that a merger of $9~{\rm M_{\odot}}$ and $8~{\rm M_{\odot}}$ stars whose asteroseismic signatures—systematically lower p-mode frequencies and distinctive g-mode period spacing variations—are clearly distinguishable from those of single stars.
Recently, \citet{Sch25} reviewed the state-of-the-art of stellar mergers in both observations and theory.

In this work, we study the asteroseismic signatures of both mass accretion and mass donation in stars with mass range of pulsating B-type ($M=3.0$–$4.5~{\rm M_{\odot}}$) and G-, F-, and A-type stars ($M=1.0$–$2.0~{\rm M_{\odot}}$). These are the favorable ranges for asteroseismology, as opposed to previous works that focused on more massive stars. We present our set-up for the stellar structure calculations with mass transfer in Section 2. In Section 3, we show the g-mode differences between accretors/donors and their single-star counterparts and link them to the internal Brunt–Väisälä and sound speed profiles and chemical gradient transition zones. Results for both B-star models (Sec. 3.1) and for lower-mass G-, F-, and A-type models (Sec. 3.2) are presented. We conclude and discuss the limitations and future prospects in Section 4.

\section{Physical inputs and model calculations}

\subsection{Physical inputs}
Stellar models were calculated using the Modules for Experiments in Stellar Astrophysics (MESA, v12778) code as developed by \citet{Pax11,Pax13,Pax15,Pax18}.  The oscillation frequencies were computed by using the stellar oscillation code GYRE \citep{Tow13,Tow18,Sun23}.

We adopt the metallicity mixture of GS98 \citep{Gre98} and the OPAL \citep{Igl96} opacity table. The Eddington gray-atmosphere $T-\tau$ relation was chosen as the stellar atmosphere model, and the convection zone was treated with the standard mixing-length theory (MLT) proposed by \citet{Cox68} with a mixing-length parameter of $\alpha_{\rm MLT}=2.0$. 
Additionally, we adopt the thermonuclear reaction net `o18\_ and \_ne22'.

Convective overshooting treatment for the convective core caused by central hydrogen burning followed the theory of \citet{Her00}. The overshooting mixing diffusion coefficient $D_{\rm ov}$ exponentially decreases from the outer boundary of the convective core based on the Schwarzschild criterion.

In this work, the overshooting parameters of the convective core were set as the recommended value $f_{\rm ov}=0.016$ and $f_{\rm ov,0}=0.001$, which is close to the asteroseismology modeling results of \citet[][SPB star KIC 10526294 is bout 3.2 $\rm M_{\odot}$]{Mor15}.
The effects of element diffusion, semiconvection, thermohaline mixing, rotation, magnetic fields, and radiative levitation were neglected in the present calculations. 
The other main physical and parameter inputs (i.e., inlist) are shown in Appendix \ref{sec_inlist}.

\subsection{Modelling binary mass transfer with single star evolution code}

\begin{figure*}
   \centering    
  \includegraphics[trim = 0 -50 0 0,width=0.7\textwidth,angle=270]{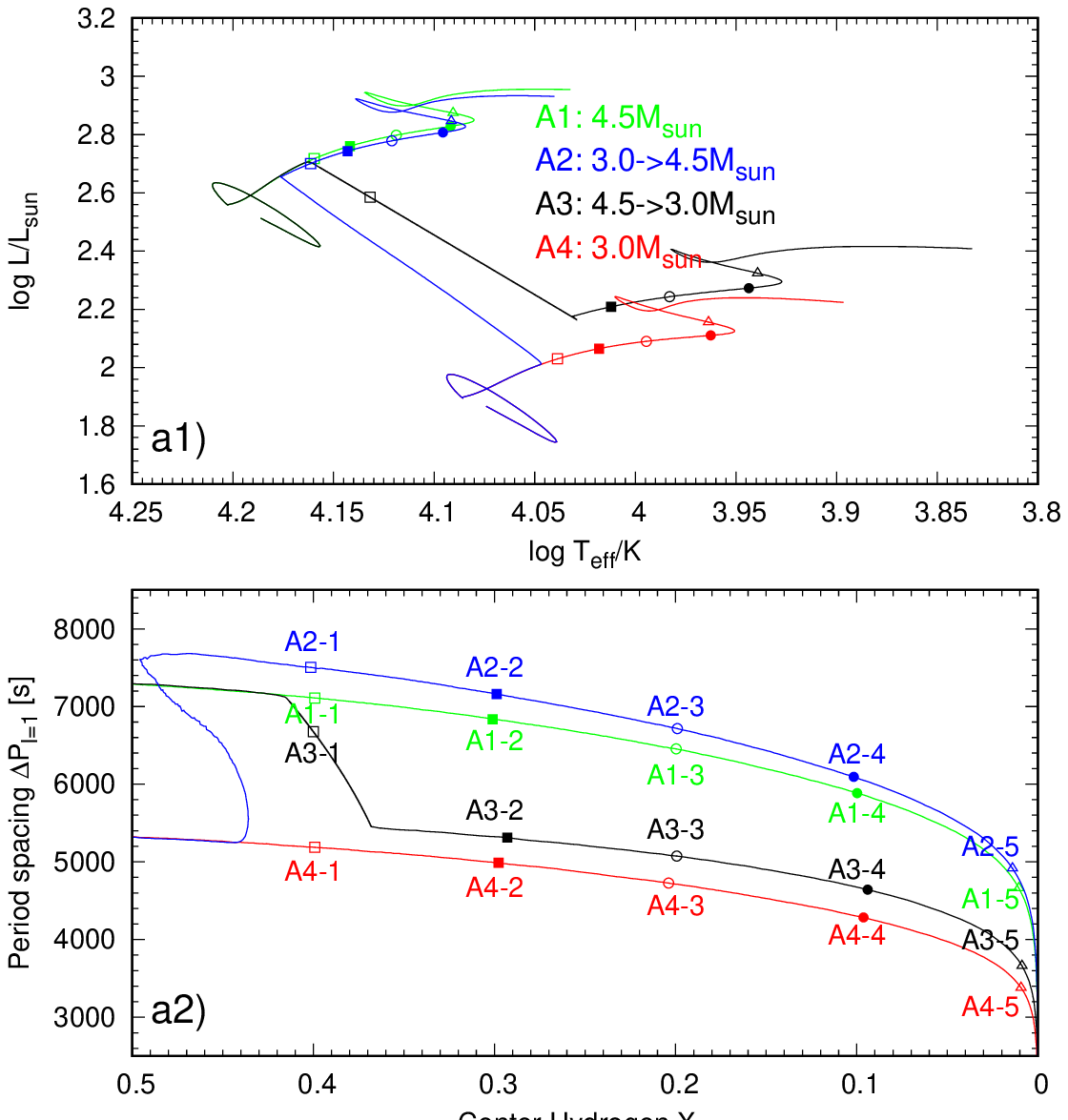}
  \quad
\includegraphics[width=0.7\textwidth,angle=270]{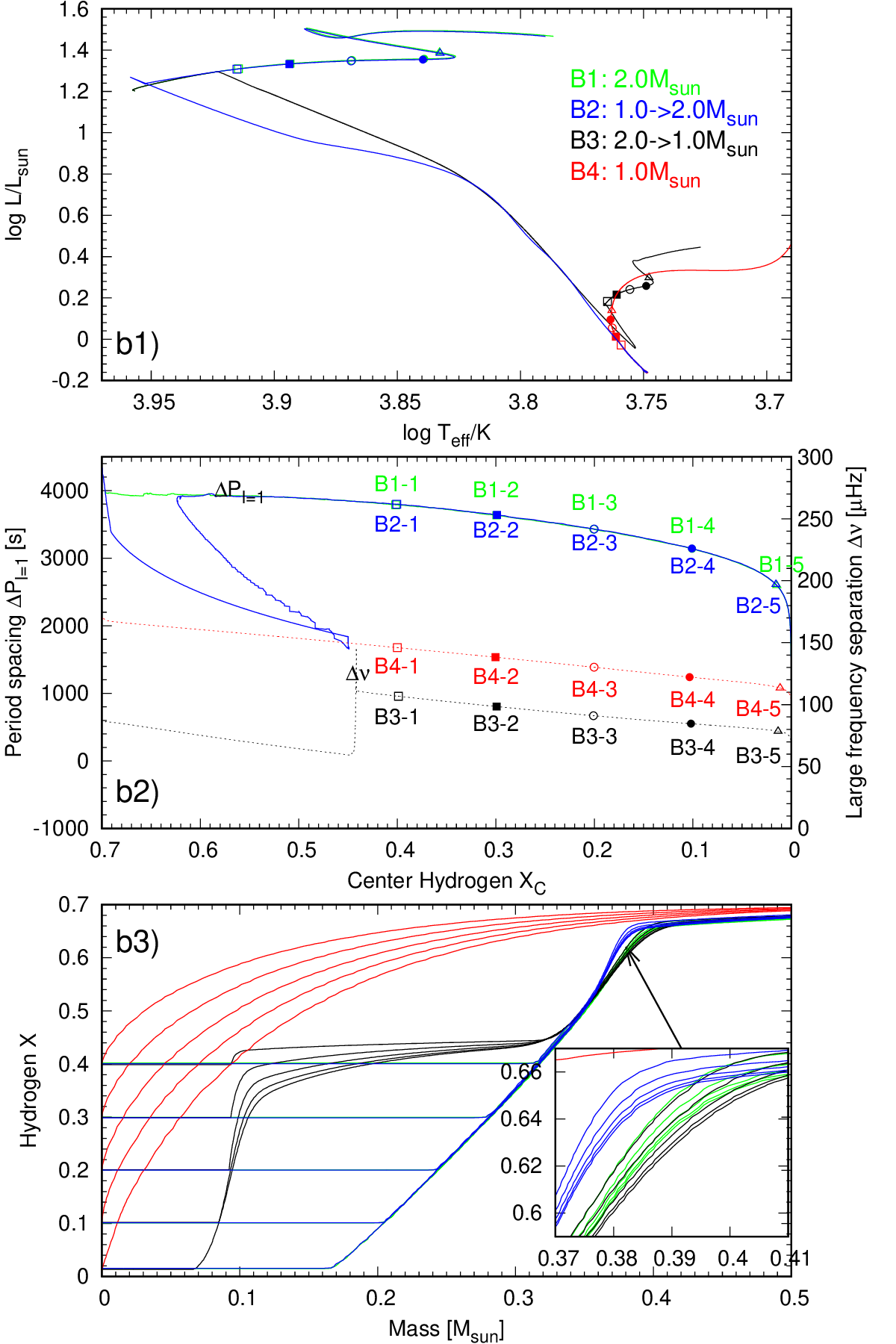}
\vspace{10pt}
   \caption{    {\it Left:} {a1)} Evolutionary tracks of four different scenarios of high-mass stars on the HR diagram.
    A1 are A4 are single-star models ($M=4.5$ and $M=3.0~{\rm M_{\odot}}$), and A2 and A3 are mass-accreting and mass-donating models ($M=3.0 \rightarrow 4.5~{\rm M_{\odot}}$ and $M=4.5\rightarrow 3.0~{\rm M_{\odot}}$). 
     {a2)} The asymptotic period spacing $\Delta P$ as a function of the central hydrogen mass fraction $X_c$. 
     {a3)} The profiles of hydrogen mass fraction $X$ for selected models are labeled by different symbols and colors in the corresponding upper and middle panels.
    {\it Right:} {b1)} Evolutionary tracks of four different scenarios of low-mass stars on the HR diagram.
     {b2)} Asymptotic period spacing $\Delta P$ as a function of $X_c$.
     {b3)} The profiles of hydrogen mass fraction $X$ in the stellar interior for selected models are labeled by different symbols and colors in b1) and b2).}
   \label{HR}%
\end{figure*}

In binary evolution, mass transfer is primarily governed by the Roche lobe geometry, which defines the critical equipotential surface surrounding each star \citep{Kop59,Egg83}. As a star evolves and its outer envelope expands, it may fill or overflow its Roche lobe, leading to the transfer of material to its companion. The onset, duration, and rate of mass transfer are determined by the system’s initial parameters—specifically the mass ratio between the donor and accretor ($q = M_{\mathrm{init,1}} / M_{\mathrm{init,2}}$), the initial orbital period ($P_{\mathrm{orb,init}}$), and the evolutionary state of the donor star \citep{Pac71,Sch24,Ceh23}. 

The mass transfer stability depends on the evolution of the Roche-lobe radius and the donor’s actual radius. Stable mass transfer generally requires mass ratio $q \leq q_{\rm crit}$, where the critical mass ratio $q_{\rm crit} \sim 1-1.5$ depending on the donor’s structure \citep{Web76, Ge15, Tau23}.

Instead of performing full binary evolution calculations, we adopt a simplified approach by imposing a prescribed mass-change rate in single-star evolution models to mimic stable mass transfer in a binary system. Although this approximation neglects some details of binary interaction, it adequately captures the effects of mass loss and accretion on the star’s internal structure and evolutionary state. The onset time, duration, and rate of mass transfer are treated as adjustable parameters, allowing us to construct reference evolutionary tracks and models with ease. This setup facilitates direct comparison with single-star (non–mass-transfer) counterparts and enables systematic follow-up analyses. Furthermore, we assume that the composition of the outer envelope remains unchanged during evolution; that is, the accreted material is assumed to have the same composition as the original stellar surface.

\subsection{Model calculations}

In our models, we set the mass change rate ``mass\_change'' to achieve mass loss when it is negative or mass gain when it is positive. For the scenario of mass loss, we set the value of ``min\_star\_massfor\_loss'' to end the process of mass loss. While, for mass gain, the set of ``max\_star\_mass\_for\_gain'' is used to end the process of mass gain. This means when stellar mass is higher or lower than ``max\_star\_mass\_for\_gain'' or ``min\_star\_mass\_for\_loss'', the process of mass gain or mass loss finishes.

We calculated two evolutionary tracks representing the B-star regime ($M=3.0,~4.5~{\rm M_{\odot}}$) which can represent SPB stars.
The other two tracks are for F, A-star regime ($M=1.0,~2.0~{\rm M_{\odot}}$), compatible with the classical $\delta$ Scuti/$\gamma$ Dor type pulsators and solar-like oscillators. Detailed descriptions are as follows,  and the corresponding evolutionary tracks are shown in Figure \ref{HR}.  

{High-mass case (B stars):} 
Stars evolve with initial masses $M_{\rm init}=3,~\text {and}~4.5~{\rm M_{\odot}}$. 

As shown in Figure \ref{HR} a1), A1 and A4 are the reference evolutionary tracks (no mass transfer, single-star), with initial masses of $4.5 ~\text{and}~3.0~\rm M_{\odot}$, respectively.

A2 is a mass-accreting track ($3.0~\rm M_{\odot} \rightarrow 4.5~\rm M_{\odot}$) with an initial mass of $M_{\rm init}=3.0~\rm M_{\odot}$, which begins mass accretion at $X_c=0.45$ with a mass-transfer rate of $\dot{M}=+6\times10^{-8}~{\rm M_{\odot}/yr}$ and the mass accretion stops at $M_{\rm F}=4.5~\rm M_{\odot}$. 

A3 is a mass-donating track ($4.5~\rm M_{\odot} \rightarrow 3.0~\rm M_{\odot}$) with an initial mass of $M_{\rm init}=4.5~\rm M_{\odot}$, and the model starts to lose mass at $X_c=0.45$ with a rate of $\dot{M}= - 6\times10^{-8}~{\rm M_{\odot}/yr}$.
The mass donation ends at $M_{\rm F}=3.0~\rm M_{\odot}$. 

{Low-mass case (A, F, G-stars):} the initial masses are $M_{\rm init}=1, ~\text {and} ~2 ~\rm M_{\odot}$.
Mass transfer is also initiated at $X_c=0.45$. 
The mass transfer rates are set to $3\times10^{-8}~{\rm M_{\odot}/yr}$. 

As shown in Figure \ref{HR} b1), B1 and B4 are the reference evolutionary tracks (no mass transfer), with $2.0,~\text{and}~1.0~\rm M_{\odot}$, respectively.

B2 is a mass-accreting track ($1.0~\rm M_{\odot} \rightarrow 2.0~\rm M_{\odot}$) with a lower initial mass of $1.0~\rm M_{\odot}$, which starts to accrete mass at $X_c=0.45$ with the mass-accretion rate of $\dot{M}=+3\times10^{-8}~{\rm M_{\odot}/yr}$ and mass accretion ceases at $M_{\rm F}=2.0~\rm M_{\odot}$.

B3 is a mass-donating track ($2.0~\rm M_{\odot} \rightarrow 1.0~\rm M_{\odot}$) with an initial mass of $2.0~\rm M_{\odot}$, which begins to lose mass at $X_c=0.45$ with the mass-change rate of $\dot{M}=-3\times10^{-8}~{\rm M_{\odot}/yr}$ and the mass transfer continues until the stellar mass reaches $M_{\rm F}=1.0~\rm M_{\odot}$.

Note that the adopted mass-transfer rate (of order $10^{-8} ~\rm M_\odot/\mathrm{yr}$ for either mass gain or loss) is consistent with a stable Roche-lobe overflow (RLOF) rate for binaries with our adopted stellar masses, as it remains below the accretor’s thermal (Kelvin–Helmholtz, KH) timescale mass-transfer limit $M/t_{\rm KH}$ \citep{Hur02, Sch24}.

The evolutionary tracks are shown in the upper panels a1) and b1) of Figure \ref{HR}. 
The asymptotic gravity-mode period spacings ($\Delta P_{l=1}$) are plotted as functions of the central hydrogen mass fraction $X_c$ in panels a2) and b2).
Note that for $M=1.0~\rm M_{\odot}$ (B4) and $M=2.0 \rightarrow 1.0~\rm M_{\odot}$ (B3) models, the pressure-mode large frequency separation $\Delta\nu$ is plotted instead of the gravity-mode period spacings ($\Delta P_{l=1}$)  (panel b2). The bottom panels a3) and b3) display the profiles of hydrogen for selected models, labeled by different symbols and colors in the corresponding upper and middle panels. 

These selected models (\#-1, \#-2, \#-3,\#-4, \#-5) correspond to the central hydrogen $X_c$ of 0.4, 0.3, 0.2, 0.1, and 0.01, respectively, where \# can be A1 to A4 or B1 to B4.
For all panels in Figure \ref{HR}, green and red denote higher and lower mass reference models (and/or tracks), and blue and black represent the models (and/or tracks) with mass donation and mass accretion, respectively.

For single-star models/tracks (A1 and A4), Figure \ref{HR} a2) shows that the asymptotic period spacing $\Delta\Pi_{l=1}\approx\Delta P_{l=1}$ is slowly decreasing as a function of central hydrogen $X_c$. And more massive stars have larger $\Delta P_{l=1}$ values.
We can clearly see the mass-donation effect in A3 ($M=4.5 \rightarrow 3.0$, black curve) from the corresponding decreasing $\Delta P$ values ($7000$ sec $\rightarrow$  $5400$ sec).
Similarly, the mass accretion increases the $\Delta P_{l=1}$ values ($5200$ sec $\rightarrow 7600$ sec) for the accretor model ($M=3.0 \rightarrow 4.5$ in blue). 

{The asymptotic period spacing values $\Delta\Pi_{l=1}$ between model A2-\# and A1-\#, and between A3-\# and A4-\# are on the order of $100$ seconds, which are much larger than the typical period resolution of {\it Kepler}. This difference can be discerned observationally if accurate stellar stellar masses and ages are known.} 

\section{Asteroseismic imprints of mass accretion and donation}\label{Sec.results}

\begin{figure*} 
\includegraphics[width=0.8\textwidth,angle=270]{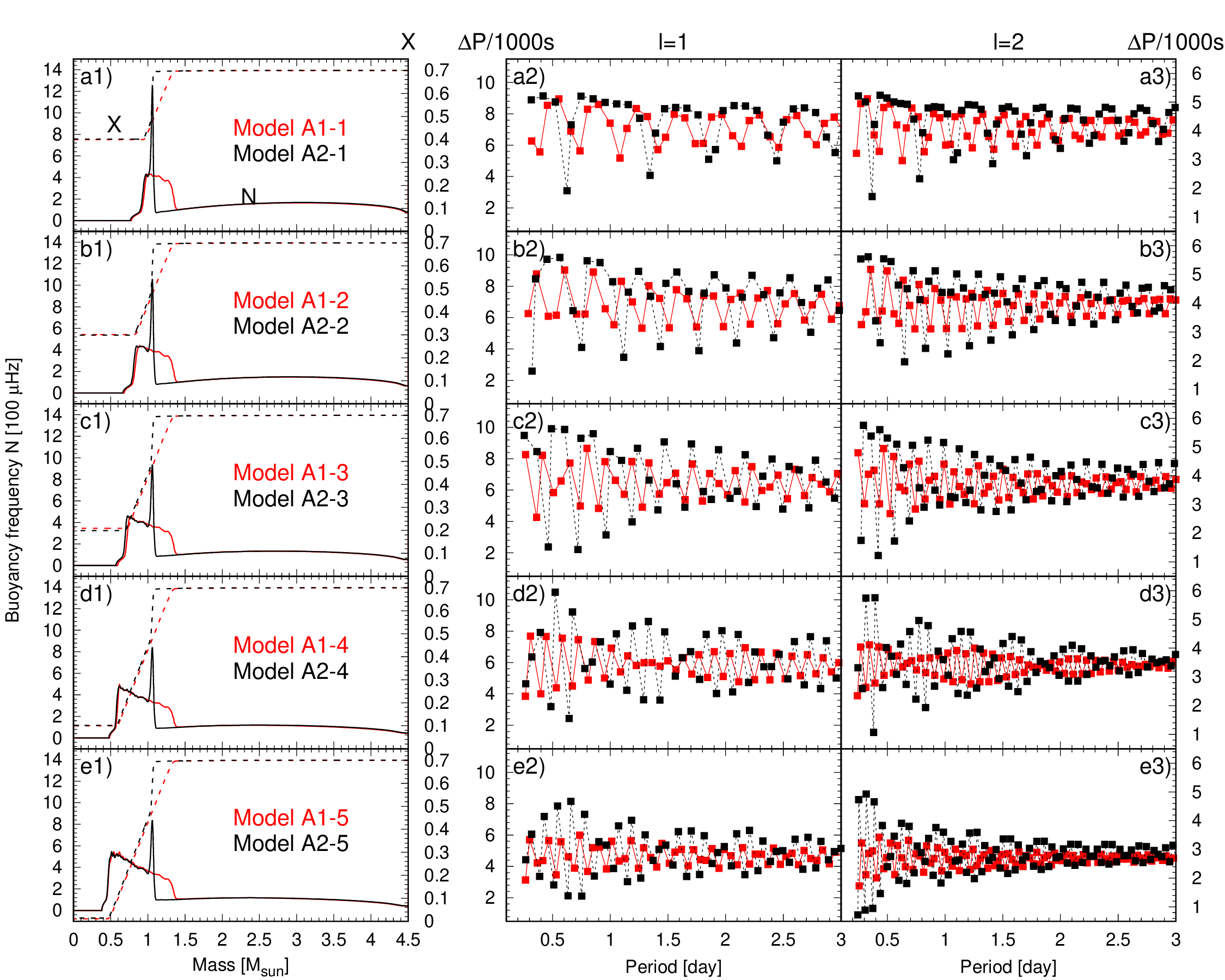}
\vspace{10pt}
    \caption{Stellar structure and oscillations of a $4.5~\rm M_{\odot}$ star from the single-star model (red) and the mass-accretion model (black).
    {Left panels:} The profiles of the buoyancy frequency $N$ (solid lines: left y-axis) and the hydrogen mass fraction $X$ (dashed lines: right y-axis). From the top to bottom, the central hydrogen $X_c$ values are 0.4, 0.3, 0.2, 0.1, 0.01.
    {Right panels:} The corresponding period spacing patterns of $l=1$ and $l=2$ g modes.}
    \label{A12}
\end{figure*}

Mass accretion or donation can significantly alter a star's internal state by changing the temperature, density, and pressure at its core \citep{Kip90, Pol94, Neo77, Ren21}.  
For main sequence stars, these changes can result in a significant fluctuation in the central thermonuclear reaction rate, which in turn  impacts the temperature gradient and convective instability \citep{Led47, Sch58}.
This is especially true for the size of the central convective core and the configuration of the chemical ($\mu$) gradient region (as shown in the bottom panels of Figure \ref{HR}).
\citet{Wag24} and \cite{Mis25b} show that mass accretion can increase the central temperatures gradient and facilitate the convective core growth (also refer to the bottom panels of Figure \ref{HR}). This is why the period spacings increase sharply when accreting mass, as shown in the blue lines in the middle panels of Figure \ref{HR}.

{These mass-transfer-induced changes can persist for an extended period until the region of thermonuclear reactions at the center moves outward into the shell beyond the center helium core and the Hydrogen shell at the outer region of the $\mu$-gradient, which is left by the early center thermonuclear reactions or mass-transfer, has been consumed and disrupts the previous state \citep{Pac71}.} {According to the theory of stellar evolution, the bottom panels of Figure \ref{HR} show that, in our calculated cases, the $\mu$-gradient imprinting in high-mass scenarios (4.0, 3.0, and 2.0 $\rm M_{\odot}$) can persist until the stars enter the horizontal branch phase. In comparison, for the low-mass case ($M = 1.0~\rm M_{\odot}$), stars will lose the signature of the $\mu$-gradient during the late stages of the red-giant branch (RGB) phase.} This means that this signal is preserved throughout most of the star’s lifetime. 
The induced asteroseismic changes will {generally be preserved} for a long time and can be detected in observations. {It is evident that the formation and persistence of the $\mu$-gradient are governed by the characteristics of the mass-transfer process, such as its onset time, rate, and duration.}

In this section, we show the asteroseismic imprints of mass accretion and donation by comparing models formed from mass transfer and models from single star evolution. We focus on g modes in upper main sequence stars ($M=3.0,~4.5,~\text{and}~2.0~\rm M_{\odot}$) and p modes in low-mass solar-like stars ($M=1.0~\rm M_{\odot}$).

\subsection{High-mass scenario}\label{Sec.results-fastmassaccretion}

\subsubsection{Mass accretion scenario: $3.0~\rm M_{\odot}$  $\rightarrow$ 4.5 $\rm M_{\odot}$}\label{Sec.results-massaccrete-massive}

In Figure \ref{A12}, we compare two sets of models with the same mass ($M=4.5~\rm M_{\odot}$) but different formation history: one from single-star evolution (red), the other from mass accretion (black).

For an upper main sequence star, as shown in Figure \ref{HR} (the red and green lines of panels a3), due to the shrinkage of the central convective core and the burning of the central hydrogen, a {gentle} $\mu$-gradient structure is formed beyond the convective core along with stellar evolution. Rapid mass accretion on the stellar surface will lead to the central temperature, density, and pressure to rapidly increase. The corresponding thermonuclear reaction rate and the corresponding region will be enlarged. This leads to a {steeper} radiative temperature gradient, which makes the region satisfying the convective instability criterion larger. 
Thus, the central convective-core boundary will expand, and enters the previous receding core along the evolution. 
The $\mu$-gradient structure will be partially or fully removed and a steep or almost discontinuous $\mu$-gradient structure (as shown in panel a3) blue lines) is formed. 

When mass accretion stops, stars return to normal evolution in a thermal time scale, and reform a new gently $\mu$-gradient structure. 
Therefore, compared to the normal evolution models, the rapid mass accretion model has narrow $\mu$-gradient regions and a steep substructure near the outer edge of the $\mu$-gradient areas, which leads to a larger buoyancy frequency $N$, as shown in panel a1)-e1) of Figure \ref{A12} in this narrow region.

The asymptotic period spacing of g modes $\Delta\Pi_l$, which is roughly the average level of observed period spacings $\Delta P$, is defined as:
\begin{equation}\label{eq_DP}
\Delta P \approx \Delta\Pi_l = \frac{\pi}{\sqrt{l(l+1)}}\Lambda_0^{-1},
\end{equation}
where $\Lambda_0$ is buoyancy radius and defined as
\begin{equation}\label{eq_lambda0}
\Lambda_0=\int_{\text{all}~N^2>0} \frac{N}{2\pi}\frac{dr}{r},
\end{equation}
which is merely related with stellar structure (with the unit of frequency, for instance, $\mu$Hz);  
$l$ is the spherical degree; $N$ is the buoyancy frequency; $r$ is the radius {coordinate}. As shown in the panel a2) of Figure \ref{HR}, the period spacings $\Delta P_{l=1}$ of the rapid mass accretion models (blue line) are larger than the single-star counterparts (green line) at the same evolutionary states.
This is because rapid mass accretion models have slight lower radius (as shown in panel a1) of Figure \ref{HR}) and a narrower $\mu$-gradient region (as shown in panels a1)-e1) of Figure \ref{A12} and the panel a3) of Figure \ref{HR}). 
Compared with the larger radius and wider $\mu$-gradient region of the normal evolution models, the effect of period spacing $\Delta P$ from the higher buoyancy frequency $N$ value in such narrow shell can be ignored (as shown in panels a1)-e1) of Figure \ref{A12}) on rapid mass accretion models. 

Similar to Equation \eqref{eq_lambda0} and the definition of the acoustic depth of p-mode, the buoyancy depth is defined as
\begin{equation}\label{eq_Lambdar}
\Lambda(r)=\int_{r}^{R} \frac{N}{2\pi}\frac{dr'}{r'}
\end{equation}
\citep[for detailed information refer to e.g.,][]{Mig08, Wu18, Wu19, Wu20}, where $R$ is the stellar radius. Similarly, we also define the buoyancy width of the $\mu$-gradient region:
\begin{equation}\label{eq_Lambdamu}
\Lambda_\mu=\int_{r_1}^{r_2} \frac{N}{2\pi}\frac{dr'}{r'}
\end{equation}
where, $r_1$ ($r_2$) is the inner (outer) boundary of the $\mu$-gradient region. For a normal intermediate- or massive single main sequence star model, $r_1$ is the boundary of the center convective core.

It can be seen from Figure \ref{A12} that the average values, varying frequency, and the varying amplitude of the period spacing patterns are affected. 
Comparing the $\Delta P$ series in the middle and right panels, we can see that the red (single star) models always have higher varying frequency and larger varying amplitude than the black (mass-accretion) models.
This is true for all evolutionary stages from the top to the bottom.

According to the analyses of \citet[][refer to their Equation (12) and Figure 5]{Wu18}, the higher varying frequency can be explained by the different width of the $\mu$-gradient region in buoyancy width $\Lambda_{\mu}$: wider buoyancy width, higher varying frequency, younger star.  
The larger period spacing varying amplitude is caused by the higher buoyancy frequency near the outer boundary of $\mu$-gradient region and by the narrower buoyancy width $\Lambda_\mu$ \citep[more details refer to the Equation (2) of][]{Wu20}. The larger period spacing varying amplitude of the mass accreting model also would mislead us to regard that the star with fast mass accretion seem to be a younger star compared to the normal single star. Therefore, the shape of the period spacings varying with period shows that the mass accreting models seem to like younger stars.

The higher varying frequency also can be explained by the $\mu$-gradient transition point being at a larger radius (or normalized buoyancy radius $u$, see below). The hydrogen mass fraction $X$ profile transition points in the left panels align with the right (outer) edge of the Brunt bump. This right edge is at much larger radius in the red models than in the black models.

The Brunt profiles in the black models also have a spike near the outer edge of the Brunt bump, which is due to the steep $\mu$-gradient.
The corresponding red models do not have this spike.
As shown by \citet{Guo25}, the Fourier amplitude spectra of period spacings is approximately equal to the buoyancy glitch derivative: 
\begin{equation}\label{eq_FT}
FT\left ( \frac{\Delta P - \Delta\Pi_l}{\Delta\Pi_l} \right )\approx \left |\frac{d(\delta N/N)}{d\ln u}\right |,
\end{equation}
where $u$ is the normalized buoyancy radius defined by:
\begin{equation}
   u(r)=1-\frac{\Lambda(r)}{\Lambda_0}=\frac{\int^r_{r_1}{N/r' dr'}} {  \int^R_{r_1}{N/r' dr'}}.
\end{equation}

The strong spike in accreting models acts as a large-amplitude buoyancy glitch, which causes g-mode period spacing variation amplitudes to be larger than the red models (middle and right panels).

Another observation is that for models of \#-4 and \#-5: A1-4 (A2-4) and A1-5 (A2-5), there is a beating pattern in the period spacing series.
This is due to the fact the sampling of $\Delta P$ in units of radial order is always 1, and if the $\Delta P$ variation frequency is close to $0.5$, an interference pattern will arise. This is similar to the Moiré pattern, seen when you take a picture of a computer screen with the cell phone camera:  the two arrays of pixels interfere and cause a beating pattern.

The right-hand side of Equation \eqref{eq_FT} is $l$ independent, this implied that the $\Delta P$ variation frequency/period is the same for $l=1$ and $l=2$ g modes, counting in the units of radial orders. This is indeed the case in all our calculations (e.g., Figures \ref{A12}, \ref{A34}, \ref{B12}).

\subsubsection{Mass donation case: $4.5  ~\rm M_{\odot} \rightarrow 3.0  ~\rm M_{\odot}$}\label{Sec.results-massloss-massive}

\begin{figure*} 
\includegraphics[width=0.8\textwidth,angle =270]{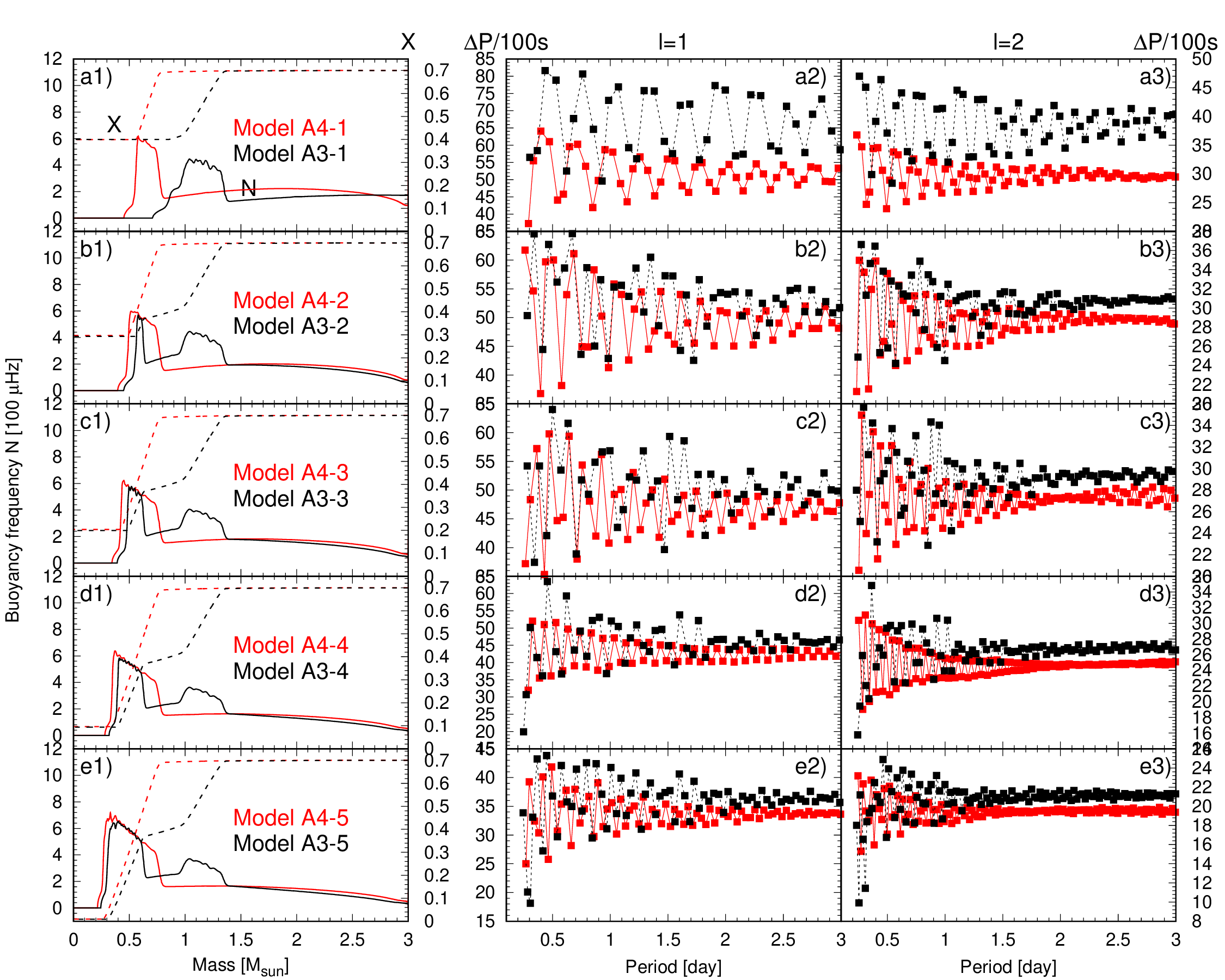}
    \vspace{10pt}
    \caption{Stellar structure and oscillations of a $3.0~{\rm M_{\odot}}$ star from the single-star model (red) and a mass-donor model (black).
    {Left panels:} The profiles of the buoyancy frequency $N$ (solid lines: left y-axis) and the hydrogen mass fraction $X$ (dashed lines: right y-axis).
    {Right panels:} The corresponding period spacing patterns of $l=1$ and $l=2$ g modes}
    \label{A34}
\end{figure*}

In Figure \ref{A34}, we compare two sets of models with the same mass ($M=3.0~\rm M_{\odot}$) but different formation history: one from single-star evolution (red), the other from mass donation (black).

As shown in the middle and right panels, the most prominent difference is the additional periodicities in the $\Delta P$ series for the black models (donor) compared to the red models (single star). This is due to the retreat of the convective core when the mass donation occurs at $X_c=0.45$ (between panel a1 and panel b1 in the left panel), which produces an additional chemical composition transition zone (two pronounced nested depletion zones, characterized by two broad troughs in the $X$ profile) and thus an additional Brunt bump at around 0.5 in mass coordinate. Two Brunt bumps are present in the A3-2, A3-3, A3-4, A3-5 models, while the corresponding single-star models in red (A4-2, A4-3, A3-4, A4-5) have only one Brunt bump.

\begin{figure*}  
\includegraphics[width=1.0\textwidth,angle =0]{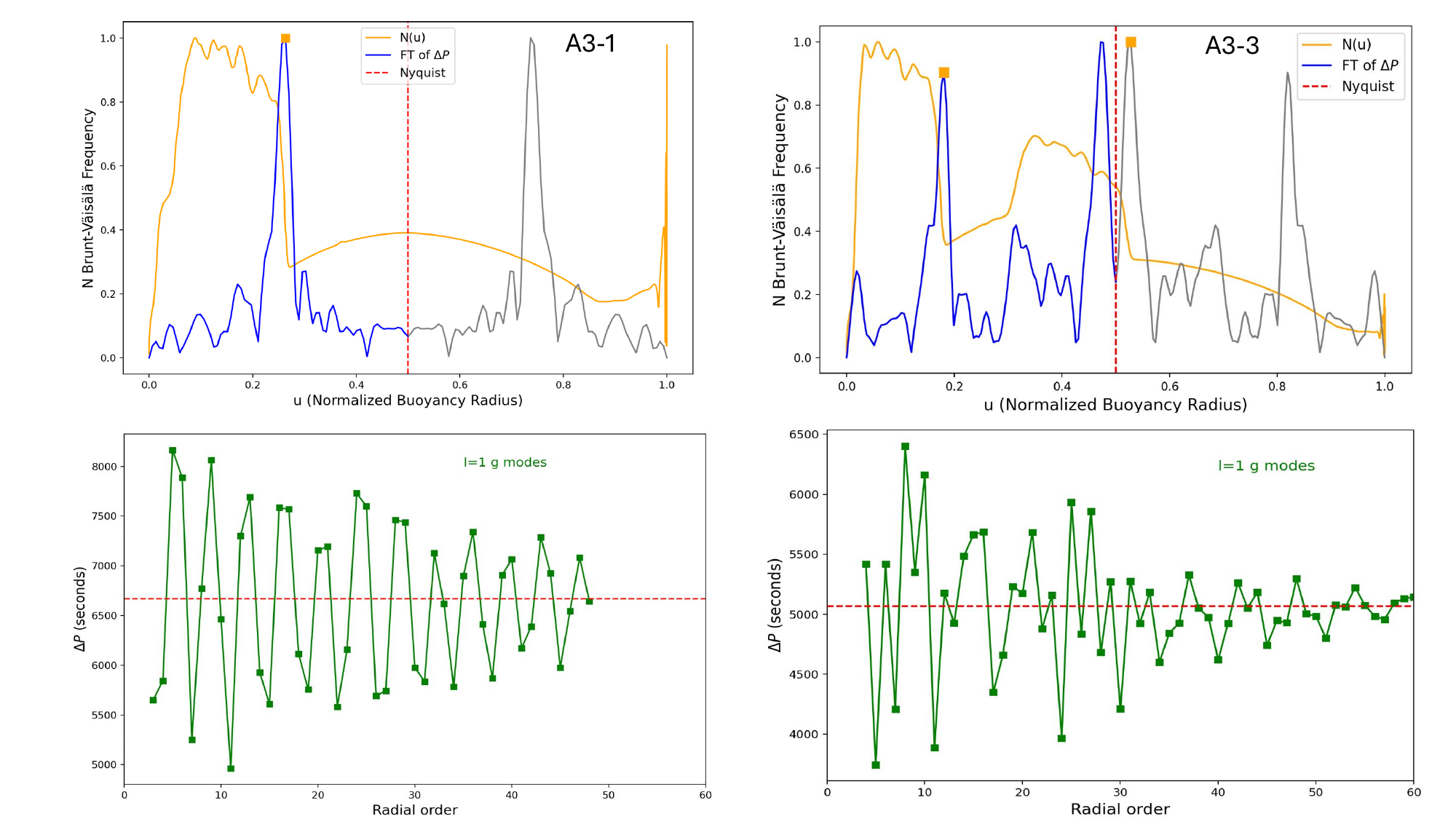}
    \vspace{10pt}
    \caption{
    {Upper panels:} Brunt–Väisälä (buoyancy) frequency profiles (orange) in two accretor models: A3-1 (left) and A3-3 (right), the later corresponds to the one with additional Brunt bumps. The normalized Fourier spectra of the period spacings of $l=1$ g modes is shown as the blue and gray lines (super Nyquist part). The vertical red dashed lines indicate the locations of the Nyquist frequency (0.5). The real peaks corresponding to the period-spacing variation are labeled by the orange squares. Note that in the Fourier spectra of A3-3 model, the dominant peak at $0.53$ is above the Nyquist frequency, and its reflection at $0.47$ is a fictitious peak.
     {Lower panels:} The corresponding period spacings of $l=1$ g modes as a function of radial order.}
    \label{fourpeak}
\end{figure*}

To quantify the periodicity in the $\Delta P$ series, we perform the Fourier transform of the $\Delta P$ with respect to radial order $n$, and show the results in Figure \ref{fourpeak}.
As can be seen, the right-edge of Brunt bump aligns with the dominant peak in the Fourier spectra of $\Delta P$ series.
For A3-1, only one dominant peak at $u \sim 0.25$ is present, corresponding to a mono-periodic variation in $\Delta P$. This is the case for both A3-1 and A4-1, although A4-1 period spacings have much lower mean values ($\Delta\Pi_l$). The real peak at $u=0.25$ labeled by the orange square is reflected to $0.75$ in the super-Nyquist (gray) part.

Starting from A3-2, two dominant peaks are present in the Fourier spectra due to the two right edges (sharp dropping) of the two Brunt bumps. As can be seen in the right panel of Figure \ref{fourpeak}, the A3-3 model g-mode period spacings have two variation frequencies, at $\sim 0.2$ and $\sim 0.48$. The gray part of the Fourier spectra is super Nyquist, and is a mirror of the sub-Nyquist ($<0.5$) part.
For mode A3-3, the real peaks are at $\sim 0.16$ and $0.53$, marked by the orange squares. These two real frequencies are reflected around $0.5$ and produce two fictitious peaks at $0.84$ and $0.47$. 
Note that the left edge of the Brunt bump also contributes to another frequency ($\sim 0.35$) in the $\Delta P$ variation, but at relatively lower amplitude.
It is worth noting that the width between the two $\mu$-gradient regions is fully determined by the duration of the mass-loss phase and the mass loss rate.

Also note the higher mean levels of the $\Delta P$ series for the black  models than the red models, due to the integration of $N/r$ (Equation \eqref{eq_DP}) in the mode propagation cavity is larger for the red models, resulting in a lower asymptotic period spacing $\Delta\Pi_l$ (also see the black line in panel a2) of Figure \ref{A12}). 

\subsection{Low mass scenario} 

\subsubsection{Mass accretion case: $1.0~\rm M_{\odot} \rightarrow 2.0~\rm M_{\odot}$}\label{Sec.results-Sec.results-massaccrete-lowmass}

\begin{figure*} 
\includegraphics[width=0.8\textwidth,angle=270]{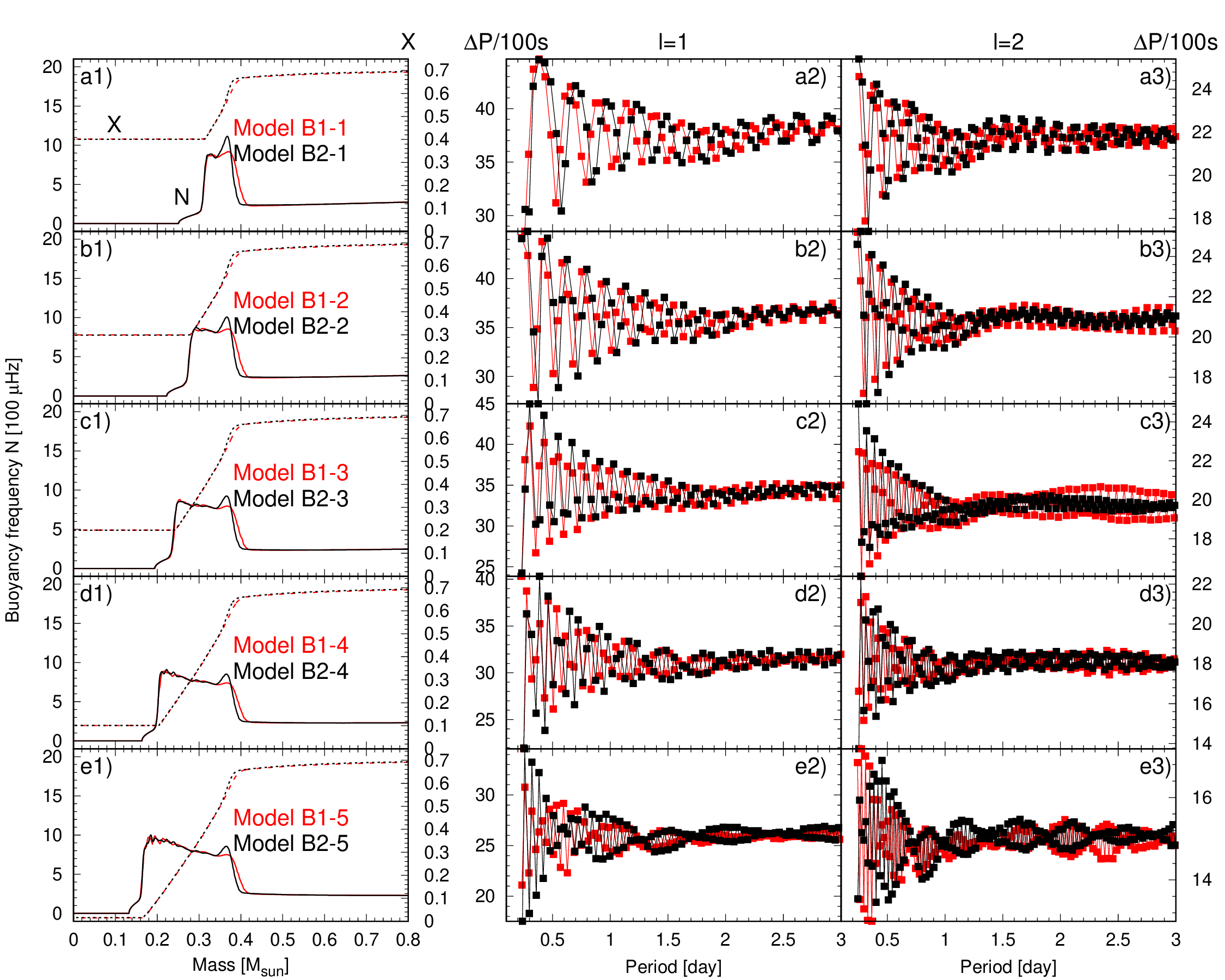}
  \vspace{10pt}
  \caption{Stellar structure and oscillations of a $2.0~\rm M_{\odot}$ star from the single-star model (red) and a mass-accretor model (black).
  {Left panels:} The profiles of the buoyancy frequency $N$ (solid lines: left y-axis) and the hydrogen mass fraction $X$ (dashed lines: right y-axis).
  {Right panels:} The corresponding period spacing patterns of $l=1$ and $l=2$ g modes}
  \label{B12}
\end{figure*}

Figure \ref{B12} illustrates how two stars of the same mass ($2.0~\rm M_{\odot}$) can differ structurally depending on their formation channel: single-star evolution (red) versus mass accretion (black).

The left panels illustrate how the Brunt-Väisälä frequency ($N$) and hydrogen mass fraction ($X$) vary within the stellar interior in the mass coordinates. Note that the $X$ profiles of the two models (red and black dotted lines) are nearly identical, almost overlapping. This is already shown in Figure \ref{HR}, panel b3), where the blue and green lines also almost coincide. The Brunt profiles are also very similar, with the red model having a slightly wider Brunt bump and a small-amplitude kink close to the right-edge of the bump. This is similar to the additional Brunt kink in the $M=3.5~\rm M_{\odot}$ accretor model calculated by \citet{Wag24} (their Figure 4).

Since the right-edge of Brunt bump is at a larger radial/mass coordinate (a larger buoyancy radius), this suggests a slightly higher frequency in the $\Delta P$ variation for red models. There is also a small phase shift between the red and black $\Delta P$ series due to the difference in the Brunt profile.
Other than these two differences, the $\Delta P$ series are very similar, they largely have the similar variation frequency and amplitude, and mean levels.

This means that it is very difficult to clearly distinguish the normal single star and the mass accretion model from their period spacing shapes for such low mass case  directly. Detailed modeling is necessary for making informed decisions.

\subsubsection{Mass donation case: $2.0~\rm M_{\odot} \rightarrow 1.0~\rm M_{\odot}$}\label{Sec.results-massloss-lowmass}

\begin{figure*}
\includegraphics[width=0.8\textwidth,angle =270]{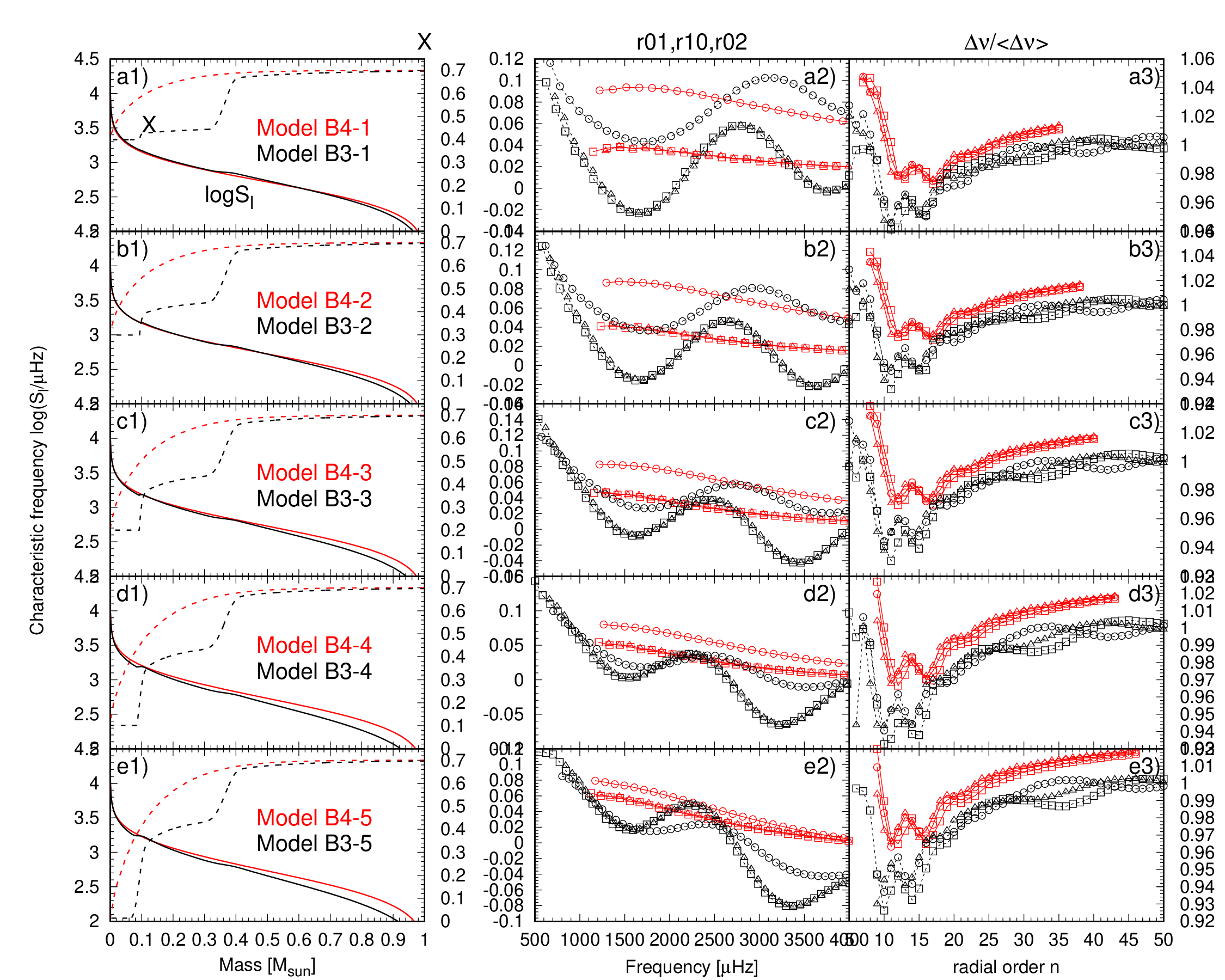}
  \vspace{10pt}
  \caption{Stellar structure and oscillations of a $1.0~\rm M_{\odot}$ star from the single-star model (red) and a mass-donor model (black).
  {Left panels:} The profiles of the Lamb frequency $\log S_{l=1}$ (solid lines: left y-axis) and the hydrogen mass fraction $X$ (dashed lines: right y-axis) in the mass coordinates (x-axis).
  {Middle panels:} The corresponding acoustic-model frequency separation ratios $r_{01},~r_{10},~\text{and}~r_{02}$ (squares,triangles, and circles) between small and large frequency separations, respectively, which are calculated with Equations \eqref{eq_ratios}.
  {Right panels:} The normalized large-frequency separation ($\Delta\nu_l/<\Delta\nu_l>$) ($l=0,~1,~2$ corresponding to squares,triangles, and circles) as a function of radial order (n). }
  \label{B34}
\end{figure*}

{It is evident that our previous analysis centered on g-mode pulsations, which are characteristic of SPB and $\gamma$ Dor pulsating variable stars. In this next phase of our analysis, we will shift our focus to p-mode pulsations, specifically solar-like oscillations. This transition is motivated by the presence of substantial outer convective envelopes in these stars, which inhibit g-modes from reaching the stellar surface and thus from being observed. In these cases, the pulsations are excited stochastically by convective turbulence, similar to the processes operating in the Sun.}   

Figure \ref{B34} contrasts two models of identical mass ($1.0~\rm M_{\odot}$) that follow distinct evolutionary paths — one formed through single-star evolution (red) and the other through mass donation (black).

The left panels show the Lamb frequency profile $S_l=\sqrt{l(l+1)}(c_s/r)=c_s k_h$ (in logarithm) and hydrogen mass fraction $X$ as functions of the enclosed mass, where $c_s$ is adiabatic sound speed and $k_h$ is the horizontal wave number of acoustic waves. 
We can immediately see the vastly different $X$ profiles for the red (single star) and black (mass-donor) models. This is also shown in panel b3) of Figure \ref{HR} (red and black lines, respectively).
Although the two stellar models have the same mass and radius, the donor model still pertains the convective core from the initial mass of $M=2.0~\rm M_{\odot}$ model. The hydrogen profiles $X(m)$ exhibit a distinct depletion region near the stellar core, forming a broad trough just outside the convective core boundary. In the more evolved models, it develops two nested depressions, reflecting successive chemical-composition gradients produced by core contraction after mass transfer.
The additional 'trough' is due to the mass donation. This is similar to the high-mass donor models case ($M=3.0~\rm M_{\odot}$, see Figure 3). 
In contrast, the single-star ($1.0~\rm M_{\odot}$) model, equivalent to the Sun, has a radiative core and a smooth $X$ profile (red dashed lines in Figure \ref{B34} left panel; red lines in Figure \ref{HR} b3). 

The middle panels of Figure \ref{B34} show the p-mode frequency separation ratios between small and large frequency separations: $r_{01}, r_{10}$ (squares and triangles) and $r_{02}$ (circles).
These frequency ratios are defined as follows \citep{Bul22}:
\begin{equation}\label{eq_ratios}
    \begin{aligned}
        r_{01}(n) &= \delta_{01}(n)/\Delta\nu_1(n) \\
        r_{10}(n) &= \delta_{10}(n)/\Delta\nu_0(n) \\
        r_{02}(n) &= \delta_{02}(n)/\Delta\nu_1(n)
    \end{aligned}
\end{equation}
The large frequency separation as a function of radial order is $\Delta\nu_l(n)=\nu_{n,l}-\nu_{n-1,l}$. The small frequency separation is $\delta_{02}(n)=\nu_{n,0} - \nu_{n-1,2}$.
The definitions of middle-point small separations $\delta_{01}~\text{and}~\delta_{10}$ {are expressed as:
\begin{equation}\label{eq_delatnu01}
   \begin{aligned}
      \delta_{01}=\frac{1}{8}(\nu_{n-1,0} -4 \nu_{n-1,1} + 6\nu_{n,0} -4\nu_{n-1,1}+\nu_{n+1,0})  \\
      \delta_{10}=-\frac{1}{8}(\nu_{n-1,1} -4 \nu_{n,0} + 6\nu_{n,1} -4\nu_{n+1,0}+\nu_{n+1,1}. ) 
   \end{aligned}
\end{equation}}
It is well known that the small separations are sensitive to the sound-speed gradient in the stellar core \citep{JCD84,Rox03,Rox09}.

In Figure \ref{B34}, the gradient of the Lamb frequency exhibits discontinuities (sign changes) at locations where the hydrogen abundance $X$ varies sharply in the mass-donor model (black). These features are most prominent at mass coordinates of approximately $0.1$ and $0.35$–$0.4$, where the slope of $\log S_l$ in the left panels changes from negative (decreasing) to positive (increasing) values.
These features can act as acoustic glitches and cause periodic variations in the small frequency separations.
The variation amplitude depends on the gradient of acoustic glitch with respect to acoustic coordinate $d(\delta c/c)/d{\ln \Phi}$ (Guo 2025b, in prep.), and the variation frequency (reciprocal of cycles measured in units of radial order) is just the local relative acoustic radius $\Phi(r_{\rm glitch})$ of the glitch \citep{Rox94,Mon03,Mig10}. $\Phi(r)$ is defined as
\begin{equation}
   \Phi(r) = \frac{ \int^r_{r_1} \frac{dr'}{c(r')}}{ \int^R_{r_1} \frac{dr'}{c(r')}}=\frac{\tau(r)}{\tau_0}
\end{equation}
where $r_1$ is the inner boundary, $R$ is the outer boundary of the acoustic wave.
As can be seen in the left panels, the red model has a smoother $X$ profile, thus smoother sound-speed (Lamb frequency) gradient, the small separation ratios do not show periodic variations (red curve in middle panels).
This is the reason for the vastly different small separation ratios (smooth red curves vs. oscillatory black curves) in the middle panels.
However, the global large frequency separations $\Delta\nu$ (right panels) are more similar for both models. This is because the variation of $\Delta\nu$ with radial order $n$ is primarily sensitive to the stellar envelope structure, rather than the core structure that differentiates the two models.

In summary, the above analyses show that the asteroseismic signals caused by mass accretion are probably too weak to be directly identified from the shape of the period-spacing series, regardless of whether it is a high-mass or low-mass scenario. Detail asteroseismic modeling might be an effective approach to distinguish mass-accretion stars from normal single stars. In contrast, for mass-donation (loss) scenarios in both high-mass and low-mass cases, the additional asteroseismic signals are prominent and they could be directly recognized from their period-spacing series or frequency-separation ratios.

\section{Discussion and conclusion}\label{Sec.discussion}

Mass accretor and donor models exhibit internal structures that are markedly different from those of single-star counterparts. 
The main-sequence evolution of intermediate- and high-mass stars naturally produces a hydrogen abundance gradient outside the convective core, which leaves a distinct chemical transition zone. In a binary system, this process can be interrupted by mass transfer. 
When mass is accreted onto a star, the central pressure and temperature increase, the nuclear energy generation rate rises, and the convective core expands. 
Conversely, in a donor undergoing mass loss, the opposite occurs: the core retreats, and the chemical gradient steepens.
These contrasting responses produce pronounced structural and asteroseismic differences between accretors, donors, and single-star models.

The sharp chemical composition transition zones (additional spikes or troughs) in the $X$ profiles, induce additional kinks or broad bumps in the Brunt profile. The corresponding g-mode period spacing series show additional periodic modulations, whose amplitude depends on the gradient of the Brunt glitch and whose frequency depends on the location of the buoyancy glitch. 
Fourier spectrum of g-mode period spacings are particularly useful because the dominant peaks' locations and amplitudes are just the $\Delta P$ variation frequency and glitch gradient amplitude, respectively.
This explains the g-mode period spacing differences in our $M=3.0,~4.5,~2.0~\rm M_{\odot}$ models.

For solar-type models, we find that a mass-donation episode can fundamentally alter the internal {{ structure}}: 
a donor remnant near $1~\rm M_\odot$ can retain a small convective core, unlike a normal solar-mass star with a radiative core. 
Moreover, the sound speed gradient profile can display multiple discontinuities, each acting as a potential acoustic glitch. 
These features modulate $p$-mode frequency separations and ratios in a measurably way. 
Such behavior, if confirmed observationally, would open a new window into the internal structure of post–mass-transfer solar-like oscillators.

Our results demonstrate that the asteroseismic signatures of mass exchange are, in principle, observable. 
The characteristic double or multiple periodicities in the $g$-mode period spacings serve as ``seismic fossils'' of a star’s binary-interaction history. 
Detecting these signatures requires a sufficient number of consecutive $g$ modes, which is now achievable thanks to continuous high-precision photometry from {TESS} \citep{TESS2015JATIS} and will be further enhanced by {PLATO} \citep{PLATO2025ExA} and {ET} \citep{GeET2022,GeET2024ChJSS,GeET2024SPIE}.
A systematic search for post–mass-transfer pulsators, combining photometric variability, binary light curves, and spectroscopic characterization, could yield the first observational constraints on historical mass-transfer rates and timings in binary systems.

The implications extend beyond individual stars. 
Binary mass exchange is common among intermediate- and massive-star populations.
In this sense, asteroseismology provides a complementary tool to discover interior evidence for rejuvenation or stripping events.

We summarize the main limitations of the present work.

(1) We have assumed that the accreted material has the same composition as the accretor’s surface. 
In reality, the transferred material may be helium-enriched or CNO-processed, leading to density inversions and thermohaline (double-diffusive) instabilities \citep{Ulr72,Sta08}. 
These processes act on a Kelvin–Helmholtz timescale $\frac{GM^2}{RL}$.
The interplay between such mixing and mode excitations remains to be quantified. Extending the current adiabatic stellar oscillation calculations to include non-adiabatic effects would provide valuable insights.

(2) We ignore the important effect of rotation on stellar oscillations. Apart from shifting the oscillation frequencies, it is also known that the fast rotation suppresses the p-mode amplitudes in $\delta$ Scuti stars and affects the mode stability of g modes and r modes in $\gamma$ Dor and SPB stars. Mass accretion naturally leads to fast rotating accretors, both the rotation and temperature affects the g modes of the accretor significantly. A similar study of the accretion effect on stellar oscillations such as \citet{Arr06} and \citet{Kum23} but for other pulsators is desirable.

(3) We use a fixed mass gain/loss rate, while in a real binary system, the mass transfer rate is determined by the orbital configuration and mass ratios \citep{Kol90}. We could perform a detailed binary star evolution with the MESA-binary. It is interesting to explore different mass-transfer rates and their effect on the stellar oscillations of the post-mass-transfer stars. The stability of mass transfer plays a crucial role in determining the final evolutionary outcomes \citep{Sob97} and the likelihood of detecting oscillations in the resulting systems.

(4) From Equation \eqref{eq_FT}, we can see the variation amplitude and frequency in the $\Delta P$ is spherical degree $l$ independent. But more information can be extracted from the mode phases. The $l=1$ and $l=2$ g modes have slightly different mode propagation cavities and turning points

(5) Other post-mass-transfer pulsators and binary populations: In this work, we focus exclusively on the effects of mass transfer on upper–main-sequence pulsators. As outlined in the Introduction, a wide variety of post–mass-transfer pulsators populate different regions of the Hertzsprung–Russell diagram. A comprehensive ``pulsator HR'' diagram for binary systems—similar to that presented by \citet{Kur22} and population synthesis like \citet{Hur02} —would be highly valuable. 
Dedicated modeling efforts are required for each class of post–mass-transfer pulsators, and such studies are already emerging, for example in recent work on blue stragglers (Briganti et al. 2025, private commu.).

Next, we will use a binary star evolution code to simulate the actual evolution of a binary system, including the processes of mass transfer, stellar rotation, and orbital evolution, while taking into account the simultaneous evolution of both stars. Most of the above deficiencies will be effectively addressed.

\begin{acknowledgements}
This work is co-supported by the National Natural Science Foundation of China (Grant No. 12288102), the B-type Strategic Priority Program of the Chinese Academy of Sciences (Grant No. XDB1160202), and the National Key R\&D Program of China (Grant No. 2021YFA1600400/2021YFA1600402). 
ZG thanks to the funding from the European Research Council (ERC) under the Horizon Europe programme (Synergy Grant agreement N◦101071505: 4D-STAR). While funded by the European Union, views and opinions expressed are however those of the author(s) only and do not necessarily reflect those of the European Union or the European Research Council. Neither the European Union nor the granting authority can be held responsible for them. ZG is also supported by STFC grant UKRI1179.
TW and YL also gratefully acknowledge the supports of NSFC of China (Grant Nos. 11973079, 12133011 and 12273104), Yunnan Fundamental Research Projects (Grant No. 202401AS070045), Youth Innovation Promotion Association of Chinese Academy of Sciences, Ten Thousand Talents Program of Yunnan for Top-notch Young Talents, the International Centre of Supernovae, Yunnan Key Laboratory (No. 202302AN360001), and China Manned Space Program (Grant No. CMS-CSST-2025-A14/A01). The authors gratefully acknowledge the computing time granted by the Yunnan Observatories, and provided on the facilities at the Yunnan Observatories Supercomputing Platform and the ``PHOENIX Supercomputing Platform'' jointly operated by the Binary Population Synthesis Group and The Stellar Astrophysics Group at Yunnan Observatories, Chinese Academy of Sciences. 
{And finally, the authors are cordially grateful to an anonymous referee for instructive advice and productive suggestions that significantly improved the quality of this paper.}
\end{acknowledgements}

\bibliographystyle{aa} 
\bibliography{accrete}

@ARTICLE{Gie13,
       author = {{Gies}, Douglas R. and {Guo}, Zhao and {Howell}, Steve B. and {Still}, Martin D. and {Boyajian}, Tabetha S. and {Hoekstra}, Abe J. and {Jek}, Kian J. and {LaCourse}, Daryll and {Winarski}, Troy},
        title = "{KIC 9406652: An Unusual Cataclysmic Variable in the Kepler Field of View}",
      journal = {\apj},
         year = 2013,
        month = sep,
       volume = {775},
       number = {1},
          eid = {64},
        pages = {64},
          doi = {10.1088/0004-637X/775/1/64},
archivePrefix = {arXiv},
       eprint = {1308.0369},
 primaryClass = {astro-ph.SR},
       adsurl = {https://ui.adsabs.harvard.edu/abs/2013ApJ...775...64G}
}

@ARTICLE{Deh22,
       author = {{Deheuvels}, S. and {Ballot}, J. and {Gehan}, C. and {Mosser}, B.},
        title = "{Seismic signature of electron degeneracy in the core of red giants: Hints for mass transfer between close red-giant companions}",
      journal = {\aap},
         year = 2022,
        month = mar,
       volume = {659},
          eid = {A106},
        pages = {A106},
          doi = {10.1051/0004-6361/202142094},
archivePrefix = {arXiv},
       eprint = {2108.11848},
 primaryClass = {astro-ph.SR},
       adsurl = {https://ui.adsabs.harvard.edu/abs/2022A&A...659A.106D}
}

@ARTICLE{Guo25,
       author = {{Guo}, Zhao},
        title = "{Asteroseismology and Buoyancy Glitch Inversion with Fourier Spectra of Gravity Mode Period Spacings}",
      journal = {arXiv e-prints},
         year = 2025,
        month = nov,
          eid = {arXiv:2511.05780},
        pages = {arXiv:2511.05780},
          doi = {10.48550/arXiv.2511.05780},
archivePrefix = {arXiv},
       eprint = {2511.05780},
 primaryClass = {astro-ph.SR},
       adsurl = {https://ui.adsabs.harvard.edu/abs/2025arXiv251105780G}
}

@ARTICLE{Arr06,
       author = {{Arras}, Phil and {Townsley}, Dean M. and {Bildsten}, Lars},
        title = "{Pulsational Instabilities in Accreting White Dwarfs}",
      journal = {\apjl},
         year = 2006,
        month = jun,
       volume = {643},
       number = {2},
        pages = {L119-L122},
          doi = {10.1086/505178},
archivePrefix = {arXiv},
       eprint = {astro-ph/0604319},
 primaryClass = {astro-ph},
       adsurl = {https://ui.adsabs.harvard.edu/abs/2006ApJ...643L.119A}
}

@ARTICLE{Bul22,
       author = {{Buldgen}, Ga{\"e}l and {B{\'e}trisey}, J{\'e}r{\^o}me and {Roxburgh}, Ian W. and {Vorontsov}, Sergei V. and {Reese}, Daniel R.},
        title = "{Inversions of Stellar Structure From Asteroseismic Data}",
      journal = {Frontiers in Astronomy and Space Sciences},
         year = 2022,
        month = jul,
       volume = {9},
          eid = {942373},
        pages = {942373},
          doi = {10.3389/fspas.2022.942373},
archivePrefix = {arXiv},
       eprint = {2206.11507},
 primaryClass = {astro-ph.SR},
       adsurl = {https://ui.adsabs.harvard.edu/abs/2022FrASS...9.2373B}
}

@ARTICLE{LiG2024AA,
       author = {{Li}, Gang and {Deheuvels}, S{\'e}bastien and {Ballot}, J{\'e}r{\^o}me},
        title = "{Asteroseismic measurement of core and envelope rotation rates for 2006 red giant branch stars}",
      journal = {\aap},
         year = 2024,
        month = aug,
       volume = {688},
          eid = {A184},
        pages = {A184},
          doi = {10.1051/0004-6361/202449882},
archivePrefix = {arXiv},
       eprint = {2405.12116},
 primaryClass = {astro-ph.SR},
       adsurl = {https://ui.adsabs.harvard.edu/abs/2024A&A...688A.184L}
}

@ARTICLE{Cor19,
       author = {{C{\'o}rsico}, Alejandro H. and {Althaus}, Leandro G. and {Miller Bertolami}, Marcelo M. and {Kepler}, S.~O.},
        title = "{Pulsating white dwarfs: new insights}",
      journal = {\aapr},
         year = 2019,
        month = sep,
       volume = {27},
       number = {1},
          eid = {7},
        pages = {7},
          doi = {10.1007/s00159-019-0118-4},
archivePrefix = {arXiv},
       eprint = {1907.00115},
 primaryClass = {astro-ph.SR},
       adsurl = {https://ui.adsabs.harvard.edu/abs/2019A&ARv..27....7C}
}

@BOOK{Cox68,
       author = {{Cox}, J.~P. and {Giuli}, R.~T.},
        title = "{Principles of stellar structure}",
         year = 1968,
       adsurl = {https://ui.adsabs.harvard.edu/abs/1968pss..book.....C}
}

@ARTICLE{Guo17,
       author = {{Guo}, Zhao and {Gies}, Douglas R. and {Matson}, Rachel A. and {Garc{\'\i}a Hern{\'a}ndez}, Antonio and {Han}, Zhanwen and {Chen}, Xuefei},
        title = "{KIC 8262223: A Post-mass Transfer Eclipsing Binary Consisting of a Delta Scuti Pulsator and a Helium White Dwarf Precursor}",
      journal = {\apj},
         year = 2017,
        month = mar,
       volume = {837},
       number = {2},
          eid = {114},
        pages = {114},
          doi = {10.3847/1538-4357/aa61a4},
archivePrefix = {arXiv},
       eprint = {1610.00350},
 primaryClass = {astro-ph.SR},
       adsurl = {https://ui.adsabs.harvard.edu/abs/2017ApJ...837..114G}
}

@ARTICLE{Guo17b,
       author = {{Guo}, Z. and {Gies}, D.~R. and {Matson}, R.~A.},
        title = "{Gravity Modes Reveal the Internal Rotation of a Post-mass-transfer Gamma Doradus/Delta Scuti Hybrid Pulsator in Kepler  Eclipsing Binary KIC 9592855}",
      journal = {\apj},
         year = 2017,
        month = dec,
       volume = {851},
       number = {1},
          eid = {39},
        pages = {39},
          doi = {10.3847/1538-4357/aa978c},
archivePrefix = {arXiv},
       eprint = {1704.03789},
 primaryClass = {astro-ph.SR},
       adsurl = {https://ui.adsabs.harvard.edu/abs/2017ApJ...851...39G}
}

@ARTICLE{Guo21,
       author = {{Guo}, Zhao},
        title = "{Asteroseismology of Close Binary Stars: Tides and Mass Transfer}",
      journal = {Frontiers in Astronomy and Space Sciences},
         year = 2021,
        month = may,
       volume = {8},
          eid = {67},
        pages = {67},
          doi = {10.3389/fspas.2021.663026},
       adsurl = {https://ui.adsabs.harvard.edu/abs/2021FrASS...8...67G}
}

@ARTICLE{Gre98,
       author = {{Grevesse}, N. and {Sauval}, A.~J.},
        title = "{Standard Solar Composition}",
      journal = {\ssr},
         year = 1998,
        month = may,
       volume = {85},
        pages = {161-174},
          doi = {10.1023/A:1005161325181},
       adsurl = {https://ui.adsabs.harvard.edu/abs/1998SSRv...85..161G}
}

@ARTICLE{Hat23,
       author = {{Hatta}, Yoshiki},
        title = "{Semi-analytical Expression of G-mode Period Spacing: The Case of Brunt-V{\"a}is{\"a}l{\"a} Frequency with Not a Jump but a Ramp}",
      journal = {\apj},
         year = 2023,
        month = jun,
       volume = {950},
       number = {2},
          eid = {165},
        pages = {165},
          doi = {10.3847/1538-4357/acd4b9},
archivePrefix = {arXiv},
       eprint = {2305.06840},
 primaryClass = {astro-ph.SR},
       adsurl = {https://ui.adsabs.harvard.edu/abs/2023ApJ...950..165H}
}

@ARTICLE{Hen24,
       author = {{Henneco}, J. and {Schneider}, F.~R.~N. and {Hekker}, S. and {Aerts}, C.},
        title = "{Merger seismology: Distinguishing massive merger products from genuine single stars using asteroseismology}",
      journal = {\aap},
         year = 2024,
        month = oct,
       volume = {690},
          eid = {A65},
        pages = {A65},
          doi = {10.1051/0004-6361/202450508},
archivePrefix = {arXiv},
       eprint = {2406.14416},
 primaryClass = {astro-ph.SR},
       adsurl = {https://ui.adsabs.harvard.edu/abs/2024A&A...690A..65H}
}

@ARTICLE{Hen25,
       author = {{Henneco}, J. and {Schneider}, F.~R.~N. and {Heller}, M. and {Hekker}, S. and {Aerts}, C.},
        title = "{Asteroseismic predictions for a massive main-sequence merger product}",
      journal = {\aap},
         year = 2025,
        month = jun,
       volume = {698},
          eid = {A49},
        pages = {A49},
          doi = {10.1051/0004-6361/202453533},
archivePrefix = {arXiv},
       eprint = {2504.08683},
 primaryClass = {astro-ph.SR},
       adsurl = {https://ui.adsabs.harvard.edu/abs/2025A&A...698A..49H}
}

@ARTICLE{Her00,
       author = {{Herwig}, F.},
        title = "{The evolution of AGB stars with convective overshoot}",
      journal = {\aap},
         year = 2000,
        month = aug,
       volume = {360},
        pages = {952-968},
          doi = {10.48550/arXiv.astro-ph/0007139},
archivePrefix = {arXiv},
       eprint = {astro-ph/0007139},
 primaryClass = {astro-ph},
       adsurl = {https://ui.adsabs.harvard.edu/abs/2000A&A...360..952H}
}

@ARTICLE{Igl96,
       author = {{Iglesias}, Carlos A. and {Rogers}, Forrest J.},
        title = "{Updated Opal Opacities}",
      journal = {\apj},
         year = 1996,
        month = jun,
       volume = {464},
        pages = {943},
          doi = {10.1086/177381},
       adsurl = {https://ui.adsabs.harvard.edu/abs/1996ApJ...464..943I}
}

@ARTICLE{Kum23,
       author = {{Kumar}, Praphull and {Townsley}, Dean M.},
        title = "{Gravity Modes on Rapidly Rotating Accreting White Dwarfs and Their Variation after Dwarf Novae}",
      journal = {\apj},
         year = 2023,
        month = jul,
       volume = {951},
       number = {2},
          eid = {122},
        pages = {122},
          doi = {10.3847/1538-4357/acd1df},
archivePrefix = {arXiv},
       eprint = {2305.03809},
 primaryClass = {astro-ph.SR},
       adsurl = {https://ui.adsabs.harvard.edu/abs/2023ApJ...951..122K}
}

@ARTICLE{Li22,
       author = {{Li}, Yaguang and {Bedding}, Timothy R. and {Murphy}, Simon J. and {Stello}, Dennis and {Chen}, Yifan and {Huber}, Daniel and {Joyce}, Meridith and {Marks}, Dion and {Zhang}, Xianfei and {Bi}, Shaolan and {Colman}, Isabel L. and {Hayden}, Michael R. and {Hey}, Daniel R. and {Li}, Gang and {Montet}, Benjamin T. and {Sharma}, Sanjib and {Wu}, Yaqian},
        title = "{Discovery of post-mass-transfer helium-burning red giants using asteroseismology}",
      journal = {Nature Astronomy},
         year = 2022,
        month = apr,
       volume = {6},
        pages = {673-680},
          doi = {10.1038/s41550-022-01648-5},
archivePrefix = {arXiv},
       eprint = {2204.06203},
 primaryClass = {astro-ph.SR},
       adsurl = {https://ui.adsabs.harvard.edu/abs/2022NatAs...6..673L}
}

@ARTICLE{Mig08,
       author = {{Miglio}, Andrea and {Montalb{\'a}n}, Josefina and {Noels}, Arlette and {Eggenberger}, Patrick},
        title = "{Probing the properties of convective cores through g modes: high-order g modes in SPB and {\ensuremath{\gamma}} Doradus stars}",
      journal = {\mnras},
         year = 2008,
        month = may,
       volume = {386},
       number = {3},
        pages = {1487-1502},
          doi = {10.1111/j.1365-2966.2008.13112.x},
archivePrefix = {arXiv},
       eprint = {0802.2057},
 primaryClass = {astro-ph},
       adsurl = {https://ui.adsabs.harvard.edu/abs/2008MNRAS.386.1487M}
}

@ARTICLE{Mig10,
       author = {{Miglio}, A. and {Montalb{\'a}n}, J. and {Carrier}, F. and {De Ridder}, J. and {Mosser}, B. and {Eggenberger}, P. and {Scuflaire}, R. and {Ventura}, P. and {D'Antona}, F. and {Noels}, A. and {Baglin}, A.},
        title = "{Evidence for a sharp structure variation inside a red-giant star}",
      journal = {\aap},
         year = 2010,
        month = sep,
       volume = {520},
          eid = {L6},
        pages = {L6},
          doi = {10.1051/0004-6361/201015442},
archivePrefix = {arXiv},
       eprint = {1009.1024},
 primaryClass = {astro-ph.SR},
       adsurl = {https://ui.adsabs.harvard.edu/abs/2010A&A...520L...6M}
}

@ARTICLE{Mkr04,
       author = {{Mkrtichian}, D.~E. and {Kusakin}, A.~V. and {Rodriguez}, E. and {Gamarova}, A. Yu. and {Kim}, C. and {Kim}, S. -L. and {Lee}, J.~W. and {Youn}, J. -H. and {Kang}, Y.~W. and {Olson}, E.~C. and {Grankin}, K.},
        title = "{Frequency spectrum of the rapidly-oscillating mass-accreting component of the Algol-type system AS Eri}",
      journal = {\aap},
         year = 2004,
        month = jun,
       volume = {419},
        pages = {1015-1024},
          doi = {10.1051/0004-6361:20040095},
       adsurl = {https://ui.adsabs.harvard.edu/abs/2004A&A...419.1015M}
}

@ARTICLE{Mkr18,
       author = {{Mkrtichian}, D.~E. and {Lehmann}, H. and {Rodr{\'\i}guez}, E. and {Olson}, E. and {Kim}, S. -L. and {Kusakin}, A.~V. and {Lee}, J.~W. and {Youn}, J. -H. and {Kwon}, S. -G. and {L{\'o}pez-Gonz{\'a}lez}, M.~J. and {Janiashvili}, E. and {Tiwari}, S.~K. and {Joshi}, Santosh and {Lampens}, P. and {Van Cauteren}, P. and {Glazunova}, L. and {Gamarova}, A. and {Grankin}, K.~N. and {Rovithis-Livaniou}, E. and {Svoboda}, P. and {Uhlar}, R. and {Tsymbal}, V. and {Kokumbaeva}, R. and {Urushadze}, T. and {Kuratov}, K. and {Shin}, H. -C. and {Kang}, Y. -W. and {Soonthornthum}, B.},
        title = "{The eclipsing binary star RZ Cas: accretion-driven variability of the multimode oscillation spectrum}",
      journal = {\mnras},
         year = 2018,
        month = apr,
       volume = {475},
       number = {4},
        pages = {4745-4767},
          doi = {10.1093/mnras/stx2841},
       adsurl = {https://ui.adsabs.harvard.edu/abs/2018MNRAS.475.4745M}
}

@ARTICLE{Mis21,
       author = {{Miszuda}, A. and {Szewczuk}, W. and {Daszy{\'n}ska-Daszkiewicz}, J.},
        title = "{The eclipsing binary systems with {\ensuremath{\delta}} Scuti component - I. KIC 10661783}",
      journal = {\mnras},
         year = 2021,
        month = aug,
       volume = {505},
       number = {3},
        pages = {3206-3218},
          doi = {10.1093/mnras/stab1597},
archivePrefix = {arXiv},
       eprint = {2105.14836},
 primaryClass = {astro-ph.SR},
       adsurl = {https://ui.adsabs.harvard.edu/abs/2021MNRAS.505.3206M}
}

@ARTICLE{Mis22,
       author = {{Miszuda}, A. and {Ko{\l}aczek-Szyma{\'n}ski}, P.~A. and {Szewczuk}, W. and {Daszy{\'n}ska-Daszkiewicz}, J.},
        title = "{The eclipsing binary systems with {\ensuremath{\delta}} Scuti component - II. AB Cas}",
      journal = {\mnras},
         year = 2022,
        month = jul,
       volume = {514},
       number = {1},
        pages = {622-639},
          doi = {10.1093/mnras/stac1197},
archivePrefix = {arXiv},
       eprint = {2204.12532},
 primaryClass = {astro-ph.SR},
       adsurl = {https://ui.adsabs.harvard.edu/abs/2022MNRAS.514..622M}
}

@ARTICLE{Mis25a,
       author = {{Miszuda}, A.},
        title = "{Off-centre convective zones in mass accreting stellar models}",
      journal = {\aap},
         year = 2025,
        month = sep,
       volume = {701},
          eid = {L7},
        pages = {L7},
          doi = {10.1051/0004-6361/202555298},
archivePrefix = {arXiv},
       eprint = {2508.14121},
 primaryClass = {astro-ph.SR},
       adsurl = {https://ui.adsabs.harvard.edu/abs/2025A&A...701L...7M}
}

@ARTICLE{Mis25b,
       author = {{Miszuda}, A. and {Guo}, Z. and {Townsend}, R.~H.~D.},
        title = "{The evolutionary and asteroseismic imprints of mass accretion. The 10 M$_\odot$ $β$ Cep case study}",
      journal = {arXiv e-prints},
         year = 2025,
        month = aug,
          eid = {arXiv:2508.19695},
        pages = {arXiv:2508.19695},
          doi = {10.48550/arXiv.2508.19695},
archivePrefix = {arXiv},
       eprint = {2508.19695},
 primaryClass = {astro-ph.SR},
       adsurl = {https://ui.adsabs.harvard.edu/abs/2025arXiv250819695M}
}

@ARTICLE{Mon03,
       author = {{Montgomery}, M.~H. and {Metcalfe}, T.~S. and {Winget}, D.~E.},
        title = "{The core/envelope symmetry in pulsating stars}",
      journal = {\mnras},
         year = 2003,
        month = sep,
       volume = {344},
       number = {2},
        pages = {657-664},
          doi = {10.1046/j.1365-8711.2003.06853.x},
archivePrefix = {arXiv},
       eprint = {astro-ph/0305601},
 primaryClass = {astro-ph},
       adsurl = {https://ui.adsabs.harvard.edu/abs/2003MNRAS.344..657M}
}

@ARTICLE{Mor15,
       author = {{Moravveji}, E. and {Aerts}, C. and {P{\'a}pics}, P.~I. and {Triana}, S.~A. and {Vandoren}, B.},
        title = "{Tight asteroseismic constraints on core overshooting and diffusive mixing in the slowly rotating pulsating B8.3V star KIC 10526294}",
      journal = {\aap},
         year = 2015,
        month = aug,
       volume = {580},
          eid = {A27},
        pages = {A27},
          doi = {10.1051/0004-6361/201425290},
archivePrefix = {arXiv},
       eprint = {1505.06902},
 primaryClass = {astro-ph.SR},
       adsurl = {https://ui.adsabs.harvard.edu/abs/2015A&A...580A..27M}
}

@ARTICLE{Neo77,
       author = {{Neo}, S. and {Miyaji}, S. and {Nomoto}, K. and {Sugimoto}, D.},
        title = "{Effect of Rapid Mass Accretion onto the Main-Sequence Stars}",
      journal = {\pasj},
         year = 1977,
        month = jan,
       volume = {29},
        pages = {249-262},
       adsurl = {https://ui.adsabs.harvard.edu/abs/1977PASJ...29..249N}
}

@ARTICLE{Pax11,
       author = {{Paxton}, Bill and {Bildsten}, Lars and {Dotter}, Aaron and {Herwig}, Falk and {Lesaffre}, Pierre and {Timmes}, Frank},
        title = "{Modules for Experiments in Stellar Astrophysics (MESA)}",
      journal = {\apjs},
         year = 2011,
        month = jan,
       volume = {192},
       number = {1},
          eid = {3},
        pages = {3},
          doi = {10.1088/0067-0049/192/1/3},
archivePrefix = {arXiv},
       eprint = {1009.1622},
 primaryClass = {astro-ph.SR},
       adsurl = {https://ui.adsabs.harvard.edu/abs/2011ApJS..192....3P}
}

@ARTICLE{Pax13,
       author = {{Paxton}, Bill and {Cantiello}, Matteo and {Arras}, Phil and {Bildsten}, Lars and {Brown}, Edward F. and {Dotter}, Aaron and {Mankovich}, Christopher and {Montgomery}, M.~H. and {Stello}, Dennis and {Timmes}, F.~X. and {Townsend}, Richard},
        title = "{Modules for Experiments in Stellar Astrophysics (MESA): Planets, Oscillations, Rotation, and Massive Stars}",
      journal = {\apjs},
         year = 2013,
        month = sep,
       volume = {208},
       number = {1},
          eid = {4},
        pages = {4},
          doi = {10.1088/0067-0049/208/1/4},
archivePrefix = {arXiv},
       eprint = {1301.0319},
 primaryClass = {astro-ph.SR},
       adsurl = {https://ui.adsabs.harvard.edu/abs/2013ApJS..208....4P}
}

@ARTICLE{Pax15,
       author = {{Paxton}, Bill and {Marchant}, Pablo and {Schwab}, Josiah and {Bauer}, Evan B. and {Bildsten}, Lars and {Cantiello}, Matteo and {Dessart}, Luc and {Farmer}, R. and {Hu}, H. and {Langer}, N. and {Townsend}, R.~H.~D. and {Townsley}, Dean M. and {Timmes}, F.~X.},
        title = "{Modules for Experiments in Stellar Astrophysics (MESA): Binaries, Pulsations, and Explosions}",
      journal = {\apjs},
         year = 2015,
        month = sep,
       volume = {220},
       number = {1},
          eid = {15},
        pages = {15},
          doi = {10.1088/0067-0049/220/1/15},
archivePrefix = {arXiv},
       eprint = {1506.03146},
 primaryClass = {astro-ph.SR},
       adsurl = {https://ui.adsabs.harvard.edu/abs/2015ApJS..220...15P}
}

@ARTICLE{Pax18,
       author = {{Paxton}, Bill and {Schwab}, Josiah and {Bauer}, Evan B. and {Bildsten}, Lars and {Blinnikov}, Sergei and {Duffell}, Paul and {Farmer}, R. and {Goldberg}, Jared A. and {Marchant}, Pablo and {Sorokina}, Elena and {Thoul}, Anne and {Townsend}, Richard H.~D. and {Timmes}, F.~X.},
        title = "{Modules for Experiments in Stellar Astrophysics (MESA): Convective Boundaries, Element Diffusion, and Massive Star Explosions}",
      journal = {\apjs},
         year = 2018,
        month = feb,
       volume = {234},
       number = {2},
          eid = {34},
        pages = {34},
          doi = {10.3847/1538-4365/aaa5a8},
archivePrefix = {arXiv},
       eprint = {1710.08424},
 primaryClass = {astro-ph.SR},
       adsurl = {https://ui.adsabs.harvard.edu/abs/2018ApJS..234...34P}
}

@ARTICLE{Rox94,
       author = {{Roxburgh}, I.~W. and {Vorontsov}, S.~V.},
        title = "{Seismology of the Solar Envelope - the Base of the Convective Zone as Seen in the Phase Shift of Acoustic Waves}",
      journal = {\mnras},
         year = 1994,
        month = jun,
       volume = {268},
        pages = {880},
          doi = {10.1093/mnras/268.4.880},
       adsurl = {https://ui.adsabs.harvard.edu/abs/1994MNRAS.268..880R}
}

@ARTICLE{Ren21,
       author = {{Renzo}, M. and {G{\"o}tberg}, Y.},
        title = "{Evolution of Accretor Stars in Massive Binaries: Broader Implications from Modeling {\ensuremath{\zeta}} Ophiuchi}",
      journal = {\apj},
         year = 2021,
        month = dec,
       volume = {923},
       number = {2},
          eid = {277},
        pages = {277},
          doi = {10.3847/1538-4357/ac29c5},
archivePrefix = {arXiv},
       eprint = {2107.10933},
 primaryClass = {astro-ph.SR},
       adsurl = {https://ui.adsabs.harvard.edu/abs/2021ApJ...923..277R}
}

@ARTICLE{Rui21,
       author = {{Rui}, Nicholas Z. and {Fuller}, Jim},
        title = "{Asteroseismic fingerprints of stellar mergers}",
      journal = {\mnras},
         year = 2021,
        month = dec,
       volume = {508},
       number = {2},
        pages = {1618-1631},
          doi = {10.1093/mnras/stab2528},
archivePrefix = {arXiv},
       eprint = {2108.10322},
 primaryClass = {astro-ph.SR},
       adsurl = {https://ui.adsabs.harvard.edu/abs/2021MNRAS.508.1618R}
}

@ARTICLE{Sch25,
       author = {{Schneider}, Fabian R.~N.},
        title = "{Theory, Simulations and Observations of Stellar Mergers}",
      journal = {arXiv e-prints},
         year = 2025,
        month = sep,
          eid = {arXiv:2509.18421},
        pages = {arXiv:2509.18421},
          doi = {10.48550/arXiv.2509.18421},
archivePrefix = {arXiv},
       eprint = {2509.18421},
 primaryClass = {astro-ph.SR},
       adsurl = {https://ui.adsabs.harvard.edu/abs/2025arXiv250918421S}
}

@ARTICLE{Sun23,
       author = {{Sun}, Meng and {Townsend}, R.~H.~D. and {Guo}, Zhao},
        title = "{gyre\_tides: Modeling Binary Tides within the GYRE Stellar Oscillation Code}",
      journal = {\apj},
         year = 2023,
        month = mar,
       volume = {945},
       number = {1},
          eid = {43},
        pages = {43},
          doi = {10.3847/1538-4357/acb33a},
archivePrefix = {arXiv},
       eprint = {2301.06599},
 primaryClass = {astro-ph.SR},
       adsurl = {https://ui.adsabs.harvard.edu/abs/2023ApJ...945...43S}
}

@ARTICLE{Tow13,
       author = {{Townsend}, R.~H.~D. and {Teitler}, S.~A.},
        title = "{GYRE: an open-source stellar oscillation code based on a new Magnus Multiple Shooting scheme}",
      journal = {\mnras},
         year = 2013,
        month = nov,
       volume = {435},
       number = {4},
        pages = {3406-3418},
          doi = {10.1093/mnras/stt1533},
archivePrefix = {arXiv},
       eprint = {1308.2965},
 primaryClass = {astro-ph.SR},
       adsurl = {https://ui.adsabs.harvard.edu/abs/2013MNRAS.435.3406T}
}

@ARTICLE{Tow18,
       author = {{Townsend}, R.~H.~D. and {Goldstein}, J. and {Zweibel}, E.~G.},
        title = "{Angular momentum transport by heat-driven g-modes in slowly pulsating B stars}",
      journal = {\mnras},
         year = 2018,
        month = mar,
       volume = {475},
       number = {1},
        pages = {879-893},
          doi = {10.1093/mnras/stx3142},
archivePrefix = {arXiv},
       eprint = {1712.02420},
 primaryClass = {astro-ph.SR},
       adsurl = {https://ui.adsabs.harvard.edu/abs/2018MNRAS.475..879T}
}

@ARTICLE{Wag24,
       author = {{Wagg}, Tom and {Johnston}, Cole and {Bellinger}, Earl P. and {Renzo}, Mathieu and {Townsend}, Richard and {de Mink}, Selma E.},
        title = "{The asteroseismic imprints of mass transfer. A case study of a binary mass-gainer in the SPB instability strip}",
      journal = {\aap},
         year = 2024,
        month = jul,
       volume = {687},
          eid = {A222},
        pages = {A222},
          doi = {10.1051/0004-6361/202449912},
archivePrefix = {arXiv},
       eprint = {2403.05627},
 primaryClass = {astro-ph.SR},
       adsurl = {https://ui.adsabs.harvard.edu/abs/2024A&A...687A.222W}
}

@ARTICLE{Web76,
       author = {{Webbink}, Ronald F.},
        title = "{The Evolution of Low-Mass Close Binary Systems. II. 1.50 M$_{sun}$ + 0.75 M$_{sun}$: Evolution Into Contact}",
      journal = {\apjs},
         year = 1976,
        month = nov,
       volume = {32},
        pages = {583},
          doi = {10.1086/190408},
       adsurl = {https://ui.adsabs.harvard.edu/abs/1976ApJS...32..583W}
}

@ARTICLE{Wu18,
       author = {{Wu}, Tao and {Li}, Yan and {Deng}, Zhen-min},
        title = "{A New C-D-like Diagram for SPB Stars: The Variations of Period Spacing as a Signature of Evolutionary Status}",
      journal = {\apj},
         year = 2018,
        month = nov,
       volume = {867},
       number = {1},
          eid = {47},
        pages = {47},
          doi = {10.3847/1538-4357/aadf85},
archivePrefix = {arXiv},
       eprint = {1809.02620},
 primaryClass = {astro-ph.SR},
       adsurl = {https://ui.adsabs.harvard.edu/abs/2018ApJ...867...47W}
}

@ARTICLE{Wu19,
       author = {{Wu}, Tao and {Li}, Yan},
        title = "{High-precision Asteroseismology in a Slowly Pulsating B Star: HD 50230}",
      journal = {\apj},
         year = 2019,
        month = aug,
       volume = {881},
       number = {1},
          eid = {86},
        pages = {86},
          doi = {10.3847/1538-4357/ab2ad8},
archivePrefix = {arXiv},
       eprint = {1907.04968},
 primaryClass = {astro-ph.SR},
       adsurl = {https://ui.adsabs.harvard.edu/abs/2019ApJ...881...86W}
}

@ARTICLE{Wu20,
       author = {{Wu}, Tao and {Li}, Yan and {Deng}, Zhen-min and {Lin}, Gui-fang and {Song}, Han-feng and {Jiang}, Chen},
        title = "{Asteroseismic Analyses of Slowly Pulsating B Star KIC 8324482: Ultraweak Element Mixing beyond the Central Convective Core}",
      journal = {\apj},
         year = 2020,
        month = aug,
       volume = {899},
       number = {1},
          eid = {38},
        pages = {38},
          doi = {10.3847/1538-4357/aba430},
       adsurl = {https://ui.adsabs.harvard.edu/abs/2020ApJ...899...38W}
}

@ARTICLE{Zha23,
       author = {{Zhang}, Qian-Sheng and {Li}, Yan and {Wu}, Tao and {Jiang}, Chen},
        title = "{Asteroseismic Investigation on KIC 10526294 to Probe Convective Core Overshoot Mixing}",
      journal = {\apj},
         year = 2023,
        month = aug,
       volume = {953},
       number = {1},
          eid = {9},
        pages = {9},
          doi = {10.3847/1538-4357/acde58},
archivePrefix = {arXiv},
       eprint = {2305.16721},
 primaryClass = {astro-ph.SR},
       adsurl = {https://ui.adsabs.harvard.edu/abs/2023ApJ...953....9Z}
}

@ARTICLE{Kip67,
       author = {{Kippenhahn}, R. and {Weigert}, A.},
        title = "{Entwicklung in engen Doppelsternsystemen I. Massenaustausch vor und nach Beendigung des zentralen Wasserstoff-Brennens}",
      journal = {\zap},
         year = 1967,
        month = jan,
       volume = {65},
        pages = {251},
       adsurl = {https://ui.adsabs.harvard.edu/abs/1967ZA.....65..251K}
}

@ARTICLE{pac71,
       author = {{Paczy{\'n}ski}, B.},
        title = "{Evolutionary Processes in Close Binary Systems}",
      journal = {\araa},
         year = 1971,
        month = jan,
       volume = {9},
        pages = {183},
          doi = {10.1146/annurev.aa.09.090171.001151},
       adsurl = {https://ui.adsabs.harvard.edu/abs/1971ARA&A...9..183P}
}

@ARTICLE{Pack81,
       author = {{Packet}, W.},
        title = "{On the spin-up of the mass accreting component in a close binary system}",
      journal = {\aap},
         year = 1981,
        month = sep,
       volume = {102},
       number = {1},
        pages = {17-19},
       adsurl = {https://ui.adsabs.harvard.edu/abs/1981A&A...102...17P}
}

@ARTICLE{Ibe93,
       author = {{Iben}, Jr., Icko and {Livio}, Mario},
        title = "{Common Envelopes in Binary Star Evolution}",
      journal = {\pasp},
         year = 1993,
        month = dec,
       volume = {105},
        pages = {1373},
          doi = {10.1086/133321},
       adsurl = {https://ui.adsabs.harvard.edu/abs/1993PASP..105.1373I}
}

@INCOLLECTION{Tau06,
       author = {{Tauris}, T.~M. and {van den Heuvel}, E.~P.~J.},
        title = "{Formation and evolution of compact stellar X-ray sources}",
    booktitle = {Compact stellar X-ray sources},
         year = 2006,
       editor = {{Lewin}, Walter H.~G. and {van der Klis}, Michiel},
       volume = {39},
        pages = {623-665},
          doi = {10.48550/arXiv.astro-ph/0303456},
       adsurl = {https://ui.adsabs.harvard.edu/abs/2006csxs.book..623T}
}

@ARTICLE{Web84,
       author = {{Webbink}, R.~F.},
        title = "{Double white dwarfs as progenitors of R Coronae Borealis stars and type I supernovae.}",
      journal = {\apj},
         year = 1984,
        month = feb,
       volume = {277},
        pages = {355-360},
          doi = {10.1086/161701},
       adsurl = {https://ui.adsabs.harvard.edu/abs/1984ApJ...277..355W}
}

@BOOK{Kip90,
       author = {{Kippenhahn}, Rudolf and {Weigert}, Alfred},
        title = "{Stellar Structure and Evolution}",
         year = 1990,
       adsurl = {https://ui.adsabs.harvard.edu/abs/1990sse..book.....K}
}

@ARTICLE{Pol94,
       author = {{Pols}, O.~R.},
        title = "{Case A evolution of massive close binaries: formation of contact systems and possible reversal of the supernova order}",
      journal = {\aap},
         year = 1994,
        month = oct,
       volume = {290},
        pages = {119-128},
       adsurl = {https://ui.adsabs.harvard.edu/abs/1994A&A...290..119P}
}

@ARTICLE{Led47,
       author = {{Ledoux}, P.},
        title = "{Stellar Models with Convection and with Discontinuity of the Mean Molecular Weight}",
      journal = {\apj},
         year = 1947,
        month = mar,
       volume = {105},
        pages = {305},
          doi = {10.1086/144905},
       adsurl = {https://ui.adsabs.harvard.edu/abs/1947ApJ...105..305L}
}

@BOOK{Sch58,
       author = {{Schwarzschild}, Martin},
        title = "{Structure and evolution of the stars.}",
         year = 1958,
       adsurl = {https://ui.adsabs.harvard.edu/abs/1958ses..book.....S}
}

@ARTICLE{Van98,
       author = {{Vanbeveren}, D. and {De Loore}, C. and {Van Rensbergen}, W.},
        title = "{Massive stars}",
      journal = {\aapr},
         year = 1998,
        month = jan,
       volume = {9},
       number = {1-2},
        pages = {63-152},
          doi = {10.1007/s001590050015},
       adsurl = {https://ui.adsabs.harvard.edu/abs/1998A&ARv...9...63V}
}

@ARTICLE{Lan12,
       author = {{Langer}, N.},
        title = "{Presupernova Evolution of Massive Single and Binary Stars}",
      journal = {\araa},
         year = 2012,
        month = sep,
       volume = {50},
        pages = {107-164},
          doi = {10.1146/annurev-astro-081811-125534},
archivePrefix = {arXiv},
       eprint = {1206.5443},
 primaryClass = {astro-ph.SR},
       adsurl = {https://ui.adsabs.harvard.edu/abs/2012ARA&A..50..107L}
}

@ARTICLE{Bon44,
       author = {{Bondi}, H. and {Hoyle}, F.},
        title = "{On the mechanism of accretion by stars}",
      journal = {\mnras},
         year = 1944,
        month = jan,
       volume = {104},
        pages = {273},
          doi = {10.1093/mnras/104.5.273},
       adsurl = {https://ui.adsabs.harvard.edu/abs/1944MNRAS.104..273B}
}

@ARTICLE{San12,
       author = {{Sana}, H. and {de Mink}, S.~E. and {de Koter}, A. and {Langer}, N. and {Evans}, C.~J. and {Gieles}, M. and {Gosset}, E. and {Izzard}, R.~G. and {Le Bouquin}, J. -B. and {Schneider}, F.~R.~N.},
        title = "{Binary Interaction Dominates the Evolution of Massive Stars}",
      journal = {Science},
         year = 2012,
        month = jul,
       volume = {337},
       number = {6093},
        pages = {444},
          doi = {10.1126/science.1223344},
archivePrefix = {arXiv},
       eprint = {1207.6397},
 primaryClass = {astro-ph.SR},
       adsurl = {https://ui.adsabs.harvard.edu/abs/2012Sci...337..444S}
}

@ARTICLE{Moe17,
       author = {{Moe}, Maxwell and {Di Stefano}, Rosanne},
        title = "{Mind Your Ps and Qs: The Interrelation between Period (P) and Mass-ratio (Q) Distributions of Binary Stars}",
      journal = {\apjs},
         year = 2017,
        month = jun,
       volume = {230},
       number = {2},
          eid = {15},
        pages = {15},
          doi = {10.3847/1538-4365/aa6fb6},
archivePrefix = {arXiv},
       eprint = {1606.05347},
 primaryClass = {astro-ph.SR},
       adsurl = {https://ui.adsabs.harvard.edu/abs/2017ApJS..230...15M}
}

@ARTICLE{Lau70,
       author = {{Lauterborn}, D.},
        title = "{Evolution with mass exchange of case C for a binary system of total mass 7 M sun.}",
      journal = {\aap},
         year = 1970,
        month = jul,
       volume = {7},
        pages = {150},
       adsurl = {https://ui.adsabs.harvard.edu/abs/1970A&A.....7..150L}
}

@ARTICLE{Smi14,
       author = {{Smith}, Nathan},
        title = "{Mass Loss: Its Effect on the Evolution and Fate of High-Mass Stars}",
      journal = {\araa},
         year = 2014,
        month = aug,
       volume = {52},
        pages = {487-528},
          doi = {10.1146/annurev-astro-081913-040025},
archivePrefix = {arXiv},
       eprint = {1402.1237},
 primaryClass = {astro-ph.SR},
       adsurl = {https://ui.adsabs.harvard.edu/abs/2014ARA&A..52..487S}
}

@ARTICLE{DeM13,
       author = {{de Mink}, S.~E. and {Langer}, N. and {Izzard}, R.~G. and {Sana}, H. and {de Koter}, A.},
        title = "{The Rotation Rates of Massive Stars: The Role of Binary Interaction through Tides, Mass Transfer, and Mergers}",
      journal = {\apj},
         year = 2013,
        month = feb,
       volume = {764},
       number = {2},
          eid = {166},
        pages = {166},
          doi = {10.1088/0004-637X/764/2/166},
archivePrefix = {arXiv},
       eprint = {1211.3742},
 primaryClass = {astro-ph.SR},
       adsurl = {https://ui.adsabs.harvard.edu/abs/2013ApJ...764..166D}
}

@ARTICLE{Han01,
       author = {{Han}, Z. and {Podsiadlowski}, Ph. and {Maxted}, P.~F.~L. and {Marsh}, T.~R. and {Ivanova}, N.},
        title = "{The origin of subdwarf B stars - I. The formation channels}",
      journal = {\mnras},
         year = 2002,
        month = oct,
       volume = {336},
       number = {2},
        pages = {449-466},
          doi = {10.1046/j.1365-8711.2002.05752.x},
archivePrefix = {arXiv},
       eprint = {astro-ph/0206130},
 primaryClass = {astro-ph},
       adsurl = {https://ui.adsabs.harvard.edu/abs/2002MNRAS.336..449H}
}

@ARTICLE{Lap20,
       author = {{Laplace}, E. and {G{\"o}tberg}, Y. and {de Mink}, S.~E. and {Justham}, S. and {Farmer}, R.},
        title = "{The expansion of stripped-envelope stars: Consequences for supernovae and gravitational-wave progenitors}",
      journal = {\aap},
         year = 2020,
        month = may,
       volume = {637},
          eid = {A6},
        pages = {A6},
          doi = {10.1051/0004-6361/201937300},
archivePrefix = {arXiv},
       eprint = {2003.01120},
 primaryClass = {astro-ph.SR},
       adsurl = {https://ui.adsabs.harvard.edu/abs/2020A&A...637A...6L}
}

@ARTICLE{Wil04,
       author = {{Willems}, B. and {Kolb}, U.},
        title = "{Detached white dwarf main-sequence star binaries}",
      journal = {\aap},
         year = 2004,
        month = jun,
       volume = {419},
        pages = {1057-1076},
          doi = {10.1051/0004-6361:20040085},
archivePrefix = {arXiv},
       eprint = {astro-ph/0403090},
 primaryClass = {astro-ph},
       adsurl = {https://ui.adsabs.harvard.edu/abs/2004A&A...419.1057W}
}

@ARTICLE{Che09,
       author = {{Chen}, Xuefei and {Han}, Zhanwen},
        title = "{Primordial binary evolution and blue stragglers}",
      journal = {\mnras},
         year = 2009,
        month = jun,
       volume = {395},
       number = {4},
        pages = {1822-1836},
          doi = {10.1111/j.1365-2966.2009.14669.x},
archivePrefix = {arXiv},
       eprint = {0903.3776},
 primaryClass = {astro-ph.SR},
       adsurl = {https://ui.adsabs.harvard.edu/abs/2009MNRAS.395.1822C}
}

@ARTICLE{Riv13,
       author = {{Rivinius}, Thomas and {Carciofi}, Alex C. and {Martayan}, Christophe},
        title = "{Classical Be stars. Rapidly rotating B stars with viscous Keplerian decretion disks}",
      journal = {\aapr},
         year = 2013,
        month = oct,
       volume = {21},
          eid = {69},
        pages = {69},
          doi = {10.1007/s00159-013-0069-0},
archivePrefix = {arXiv},
       eprint = {1310.3962},
 primaryClass = {astro-ph.SR},
       adsurl = {https://ui.adsabs.harvard.edu/abs/2013A&ARv..21...69R}
}

@ARTICLE{Str18,
       author = {{Streamer}, Margaret and {Ireland}, Michael J. and {Murphy}, Simon J. and {Bento}, Joao},
        title = "{A window into {\ensuremath{\delta}} Sct stellar interiors: understanding the eclipsing binary system TT Hor}",
      journal = {\mnras},
         year = 2018,
        month = oct,
       volume = {480},
       number = {1},
        pages = {1372-1383},
          doi = {10.1093/mnras/sty1881},
archivePrefix = {arXiv},
       eprint = {1807.03917},
 primaryClass = {astro-ph.SR},
       adsurl = {https://ui.adsabs.harvard.edu/abs/2018MNRAS.480.1372S}
}

@ARTICLE{Per12,
       author = {{Perets}, Hagai B. and {{\v{S}}ubr}, Ladislav},
        title = "{The Properties of Dynamically Ejected Runaway and Hyper-runaway Stars}",
      journal = {\apj},
         year = 2012,
        month = jun,
       volume = {751},
       number = {2},
          eid = {133},
        pages = {133},
          doi = {10.1088/0004-637X/751/2/133},
archivePrefix = {arXiv},
       eprint = {1202.2356},
 primaryClass = {astro-ph.SR},
       adsurl = {https://ui.adsabs.harvard.edu/abs/2012ApJ...751..133P}
}

@BOOK{Aer10,
       author = {{Aerts}, Conny and {Christensen-Dalsgaard}, J{\o}rgen and {Kurtz}, Donald W.},
        title = "{Asteroseismology}",
         year = 2010,
          doi = {10.1007/978-1-4020-5803-5},
       adsurl = {https://ui.adsabs.harvard.edu/abs/2010aste.book.....A}
}

@BOOK{Bas17,
       author = {{Basu}, Sarbani and {Chaplin}, William J.},
        title = "{Asteroseismic Data Analysis: Foundations and Techniques}",
         year = 2017,
       adsurl = {https://ui.adsabs.harvard.edu/abs/2017asda.book.....B}
}

@ARTICLE{Kol90,
       author = {{Kolb}, U. and {Ritter}, H.},
        title = "{A comparative study of the evolution of a close binary using a standard and an improved technique for computing mass transfer.}",
      journal = {\aap},
         year = 1990,
        month = sep,
       volume = {236},
        pages = {385-392},
       adsurl = {https://ui.adsabs.harvard.edu/abs/1990A&A...236..385K}
}

@ARTICLE{Ge15,
       author = {{Ge}, Hongwei and {Webbink}, Ronald F. and {Chen}, Xuefei and {Han}, Zhanwen},
        title = "{Adiabatic Mass Loss in Binary Stars. II. From Zero-age Main Sequence to the Base of the Giant Branch}",
      journal = {\apj},
         year = 2015,
        month = oct,
       volume = {812},
       number = {1},
          eid = {40},
        pages = {40},
          doi = {10.1088/0004-637X/812/1/40},
archivePrefix = {arXiv},
       eprint = {1507.04843},
 primaryClass = {astro-ph.SR},
       adsurl = {https://ui.adsabs.harvard.edu/abs/2015ApJ...812...40G}
}

@ARTICLE{Ulr72,
       author = {{Ulrich}, Roger K.},
        title = "{Thermohaline Convection in Stellar Interiors.}",
      journal = {\apj},
         year = 1972,
        month = feb,
       volume = {172},
        pages = {165},
          doi = {10.1086/151336},
       adsurl = {https://ui.adsabs.harvard.edu/abs/1972ApJ...172..165U}
}

@ARTICLE{Sta08,
       author = {{Stancliffe}, Richard J. and {Glebbeek}, Evert},
        title = "{Thermohaline mixing and gravitational settling in carbon-enhanced metal-poor stars}",
      journal = {\mnras},
         year = 2008,
        month = oct,
       volume = {389},
       number = {4},
        pages = {1828-1838},
          doi = {10.1111/j.1365-2966.2008.13700.x},
archivePrefix = {arXiv},
       eprint = {0807.1758},
 primaryClass = {astro-ph},
       adsurl = {https://ui.adsabs.harvard.edu/abs/2008MNRAS.389.1828S}
}

@ARTICLE{Che17,
       author = {{Chen}, Xuefei and {Maxted}, P.~F.~L. and {Li}, Jiao and {Han}, Zhanwen},
        title = "{The Formation of EL CVn-type Binaries}",
      journal = {\mnras},
         year = 2017,
        month = may,
       volume = {467},
       number = {2},
        pages = {1874-1889},
          doi = {10.1093/mnras/stx115},
archivePrefix = {arXiv},
       eprint = {1604.01956},
 primaryClass = {astro-ph.SR},
       adsurl = {https://ui.adsabs.harvard.edu/abs/2017MNRAS.467.1874C}
}

@ARTICLE{Max13,
       author = {{Maxted}, Pierre F.~L. and {Serenelli}, Aldo M. and {Miglio}, Andrea and {Marsh}, Thomas R. and {Heber}, Ulrich and {Dhillon}, Vikram S. and {Littlefair}, Stuart and {Copperwheat}, Chris and {Smalley}, Barry and {Breedt}, Elm{\'e} and {Schaffenroth}, Veronika},
        title = "{Multi-periodic pulsations of a stripped red-giant star in an eclipsing binary system}",
      journal = {\nat},
         year = 2013,
        month = jun,
       volume = {498},
       number = {7455},
        pages = {463-465},
          doi = {10.1038/nature12192},
archivePrefix = {arXiv},
       eprint = {1307.1654},
 primaryClass = {astro-ph.SR},
       adsurl = {https://ui.adsabs.harvard.edu/abs/2013Natur.498..463M}
}

@ARTICLE{Gia16,
       author = {{Gianninas}, A. and {Curd}, Brandon and {Fontaine}, G. and {Brown}, Warren R. and {Kilic}, Mukremin},
        title = "{Discovery of Three Pulsating, Mixed-atmosphere, Extremely Low-mass White Dwarf Precursors}",
      journal = {\apjl},
         year = 2016,
        month = may,
       volume = {822},
       number = {2},
          eid = {L27},
        pages = {L27},
          doi = {10.3847/2041-8205/822/2/L27},
archivePrefix = {arXiv},
       eprint = {1604.04621},
 primaryClass = {astro-ph.SR},
       adsurl = {https://ui.adsabs.harvard.edu/abs/2016ApJ...822L..27G}
}

@ARTICLE{Ist16,
       author = {{Istrate}, A.~G. and {Fontaine}, G. and {Gianninas}, A. and {Grassitelli}, L. and {Marchant}, P. and {Tauris}, T.~M. and {Langer}, N.},
        title = "{Asteroseismic test of rotational mixing in low-mass white dwarfs}",
      journal = {\aap},
         year = 2016,
        month = nov,
       volume = {595},
          eid = {L12},
        pages = {L12},
          doi = {10.1051/0004-6361/201629876},
archivePrefix = {arXiv},
       eprint = {1610.08513},
 primaryClass = {astro-ph.SR},
       adsurl = {https://ui.adsabs.harvard.edu/abs/2016A&A...595L..12I}
}

@ARTICLE{Ist16a,
       author = {{Istrate}, A.~G. and {Marchant}, P. and {Tauris}, T.~M. and {Langer}, N. and {Stancliffe}, R.~J. and {Grassitelli}, L.},
        title = "{Models of low-mass helium white dwarfs including gravitational settling, thermal and chemical diffusion, and rotational mixing}",
      journal = {\aap},
         year = 2016,
        month = oct,
       volume = {595},
          eid = {A35},
        pages = {A35},
          doi = {10.1051/0004-6361/201628874},
archivePrefix = {arXiv},
       eprint = {1606.04947},
 primaryClass = {astro-ph.SR},
       adsurl = {https://ui.adsabs.harvard.edu/abs/2016A&A...595A..35I}
}

@ARTICLE{Gau17,
       author = {{Gautschy}, Alfred and {Saio}, Hideyuki},
        title = "{On binary channels to anomalous Cepheids}",
      journal = {\mnras},
         year = 2017,
        month = jul,
       volume = {468},
       number = {4},
        pages = {4419-4428},
          doi = {10.1093/mnras/stx811},
archivePrefix = {arXiv},
       eprint = {1704.00124},
 primaryClass = {astro-ph.SR},
       adsurl = {https://ui.adsabs.harvard.edu/abs/2017MNRAS.468.4419G}
}

@ARTICLE{Kar17,
       author = {{Karczmarek}, P. and {Wiktorowicz}, G. and {I{\l}kiewicz}, K. and {Smolec}, R. and {St{\k{e}}pie{\'n}}, K. and {Pietrzy{\'n}ski}, G. and {Gieren}, W. and {Belczynski}, K.},
        title = "{The occurrence of binary evolution pulsators in classical instability strip of RR Lyrae and Cepheid variables}",
      journal = {\mnras},
         year = 2017,
        month = apr,
       volume = {466},
       number = {3},
        pages = {2842-2854},
          doi = {10.1093/mnras/stw3286},
archivePrefix = {arXiv},
       eprint = {1612.00465},
 primaryClass = {astro-ph.SR},
       adsurl = {https://ui.adsabs.harvard.edu/abs/2017MNRAS.466.2842K}
}

@ARTICLE{Pie12,
       author = {{Pietrzy{\'n}ski}, G. and {Thompson}, I.~B. and {Gieren}, W. and {Graczyk}, D. and {St{\k{e}}pie{\'n}}, K. and {Bono}, G. and {Moroni}, P.~G. Prada and {Pilecki}, B. and {Udalski}, A. and {Soszy{\'n}ski}, I. and {Preston}, G.~W. and {Nardetto}, N. and {McWilliam}, A. and {Roederer}, I.~U. and {G{\'o}rski}, M. and {Konorski}, P. and {Storm}, J.},
        title = "{RR-Lyrae-type pulsations from a 0.26-solar-mass star in a binary system}",
      journal = {\nat},
         year = 2012,
        month = apr,
       volume = {484},
       number = {7392},
        pages = {75-77},
          doi = {10.1038/nature10966},
archivePrefix = {arXiv},
       eprint = {1204.1872},
 primaryClass = {astro-ph.SR},
       adsurl = {https://ui.adsabs.harvard.edu/abs/2012Natur.484...75P}
}

@ARTICLE{Pil17,
       author = {{Pilecki}, Bogumi{\l} and {Gieren}, Wolfgang and {Smolec}, Rados{\l}aw and {Pietrzy{\'n}ski}, Grzegorz and {Thompson}, Ian B. and {Anderson}, Richard I. and {Bono}, Giuseppe and {Soszy{\'n}ski}, Igor and {Kervella}, Pierre and {Nardetto}, Nicolas and {Taormina}, M{\'o}nica and {St{\c{e}}pie{\'n}}, Kazimierz and {Wielg{\'o}rski}, Piotr},
        title = "{Mass and p-factor of the Type II Cepheid OGLE-LMC-T2CEP-098 in a Binary System}",
      journal = {\apj},
         year = 2017,
        month = jun,
       volume = {842},
       number = {2},
          eid = {110},
        pages = {110},
          doi = {10.3847/1538-4357/aa6ff7},
archivePrefix = {arXiv},
       eprint = {1704.07782},
 primaryClass = {astro-ph.SR},
       adsurl = {https://ui.adsabs.harvard.edu/abs/2017ApJ...842..110P}
}

@ARTICLE{Sou25,
       author = {{Southworth}, John and {Bowman}, Dominic},
        title = "{Pulsations in Binary Star Systems}",
      journal = {arXiv e-prints},
         year = 2025,
        month = sep,
          eid = {arXiv:2509.08426},
        pages = {arXiv:2509.08426},
          doi = {10.48550/arXiv.2509.08426},
archivePrefix = {arXiv},
       eprint = {2509.08426},
 primaryClass = {astro-ph.SR},
       adsurl = {https://ui.adsabs.harvard.edu/abs/2025arXiv250908426S}
}

@BOOK{Tau23,
       author = {{Tauris}, Thomas M. and {van den Heuvel}, Edward P.~J.},
        title = "{Physics of Binary Star Evolution. From Stars to X-ray Binaries and Gravitational Wave Sources}",
         year = 2023,
          doi = {10.48550/arXiv.2305.09388},
       adsurl = {https://ui.adsabs.harvard.edu/abs/2023pbse.book.....T}
}

@ARTICLE{Kur22,
       author = {{Kurtz}, Donald W.},
        title = "{Asteroseismology Across the Hertzsprung-Russell Diagram}",
      journal = {\araa},
         year = 2022,
        month = aug,
       volume = {60},
        pages = {31-71},
          doi = {10.1146/annurev-astro-052920-094232},
       adsurl = {https://ui.adsabs.harvard.edu/abs/2022ARA&A..60...31K}
}

@BOOK{Kop59,
       author = {{Kopal}, Zdenek},
        title = "{Close binary systems}",
         year = 1959,
       adsurl = {https://ui.adsabs.harvard.edu/abs/1959cbs..book.....K}
}

@ARTICLE{Egg83,
       author = {{Eggleton}, P.~P.},
        title = "{Aproximations to the radii of Roche lobes.}",
      journal = {\apj},
         year = 1983,
        month = may,
       volume = {268},
        pages = {368-369},
          doi = {10.1086/160960},
       adsurl = {https://ui.adsabs.harvard.edu/abs/1983ApJ...268..368E}
}

@ARTICLE{Sch24,
       author = {{Sch{\"u}rmann}, C. and {Langer}, N.},
        title = "{Exploring the boundary between stable mass transfer and L$_{2}$ overflow in close binary evolution}",
      journal = {\aap},
         year = 2024,
        month = nov,
       volume = {691},
          eid = {A174},
        pages = {A174},
          doi = {10.1051/0004-6361/202450354},
archivePrefix = {arXiv},
       eprint = {2404.08615},
 primaryClass = {astro-ph.SR},
       adsurl = {https://ui.adsabs.harvard.edu/abs/2024A&A...691A.174S}
}

@ARTICLE{Ceh23,
       author = {{Cehula}, Jakub and {Pejcha}, Ond{\v{r}}ej},
        title = "{A theory of mass transfer in binary stars}",
      journal = {\mnras},
         year = 2023,
        month = sep,
       volume = {524},
       number = {1},
        pages = {471-490},
          doi = {10.1093/mnras/stad1862},
archivePrefix = {arXiv},
       eprint = {2303.05526},
 primaryClass = {astro-ph.SR},
       adsurl = {https://ui.adsabs.harvard.edu/abs/2023MNRAS.524..471C}
}

@ARTICLE{Hur02,
       author = {{Hurley}, Jarrod R. and {Tout}, Christopher A. and {Pols}, Onno R.},
        title = "{Evolution of binary stars and the effect of tides on binary populations}",
      journal = {\mnras},
         year = 2002,
        month = feb,
       volume = {329},
       number = {4},
        pages = {897-928},
          doi = {10.1046/j.1365-8711.2002.05038.x},
archivePrefix = {arXiv},
       eprint = {astro-ph/0201220},
 primaryClass = {astro-ph},
       adsurl = {https://ui.adsabs.harvard.edu/abs/2002MNRAS.329..897H}
}

@ARTICLE{Rox03,
       author = {{Roxburgh}, I.~W. and {Vorontsov}, S.~V.},
        title = "{The ratio of small to large separations of acoustic oscillations as a diagnostic of the interior of solar-like stars}",
      journal = {\aap},
         year = 2003,
        month = nov,
       volume = {411},
        pages = {215-220},
          doi = {10.1051/0004-6361:20031318},
       adsurl = {https://ui.adsabs.harvard.edu/abs/2003A&A...411..215R}
}

@ARTICLE{Rox09,
       author = {{Roxburgh}, I.~W.},
        title = "{Small separations and phase shift differences of {\ensuremath{\ell}} = 0, 1 p-modes}",
      journal = {\aap},
         year = 2009,
        month = jan,
       volume = {493},
       number = {1},
        pages = {185-191},
          doi = {10.1051/0004-6361:200811047},
       adsurl = {https://ui.adsabs.harvard.edu/abs/2009A&A...493..185R}
}

@ARTICLE{Sob97,
       author = {{Soberman}, G.~E. and {Phinney}, E.~S. and {van den Heuvel}, E.~P.~J.},
        title = "{Stability criteria for mass transfer in binary stellar evolution.}",
      journal = {\aap},
         year = 1997,
        month = nov,
       volume = {327},
        pages = {620-635},
          doi = {10.48550/arXiv.astro-ph/9703016},
archivePrefix = {arXiv},
       eprint = {astro-ph/9703016},
 primaryClass = {astro-ph},
       adsurl = {https://ui.adsabs.harvard.edu/abs/1997A&A...327..620S}
}

@ARTICLE{Rap83,
       author = {{Rappaport}, S. and {Verbunt}, F. and {Joss}, P.~C.},
        title = "{A new technique for calculations of binary stellar evolution application to magnetic braking.}",
      journal = {\apj},
         year = 1983,
        month = dec,
       volume = {275},
        pages = {713-731},
          doi = {10.1086/161569},
       adsurl = {https://ui.adsabs.harvard.edu/abs/1983ApJ...275..713R}
}

@ARTICLE{Bel02,
       author = {{Belczynski}, Krzysztof and {Kalogera}, Vassiliki and {Bulik}, Tomasz},
        title = "{A Comprehensive Study of Binary Compact Objects as Gravitational Wave Sources: Evolutionary Channels, Rates, and Physical Properties}",
      journal = {\apj},
         year = 2002,
        month = jun,
       volume = {572},
       number = {1},
        pages = {407-431},
          doi = {10.1086/340304},
archivePrefix = {arXiv},
       eprint = {astro-ph/0111452},
 primaryClass = {astro-ph},
       adsurl = {https://ui.adsabs.harvard.edu/abs/2002ApJ...572..407B}
}

@ARTICLE{Han95,
       author = {{Han}, Zhanwen and {Podsiadlowski}, Philipp and {Eggleton}, Peter P.},
        title = "{The formation of bipolar planetary nebulae and close white dwarf binaries}",
      journal = {\mnras},
         year = 1995,
        month = feb,
       volume = {272},
       number = {4},
        pages = {800-820},
          doi = {10.1093/mnras/272.4.800},
       adsurl = {https://ui.adsabs.harvard.edu/abs/1995MNRAS.272..800H}
}

@INPROCEEDINGS{JCD84,
       author = {{Christensen-Dalsgaard}, J.},
        title = "{What Will Asteroseismology Teach us}",
    booktitle = {Space Research in Stellar Activity and Variability},
         year = 1984,
       editor = {{Mangeney}, A. and {Praderie}, F.},
        month = jan,
        pages = {11},
       adsurl = {https://ui.adsabs.harvard.edu/abs/1984srps.conf...11C}
}

@ARTICLE{GeET2022,
       author = {{Ge}, Jian and {Zhang}, Hui and {Zang}, Weicheng and {Deng}, Hongping and {Mao}, Shude and {Xie}, Ji-Wei and {Liu}, Hui-Gen and {Zhou}, Ji-Lin and {Willis}, Kevin and {Huang}, Chelsea and {Howell}, Steve B. and {Feng}, Fabo and {Zhu}, Jiapeng and {Yao}, Xinyu and {Liu}, Beibei and {Aizawa}, Masataka and {Zhu}, Wei and {Li}, Ya-Ping and {Ma}, Bo and {Ye}, Quanzhi and {Yu}, Jie and {Xiang}, Maosheng and {Yu}, Cong and {Liu}, Shangfei and {Yang}, Ming and {Wang}, Mu-Tian and {Shi}, Xian and {Fang}, Tong and {Zong}, Weikai and {Liu}, Jinzhong and {Zhang}, Yu and {Zhang}, Liyun and {El-Badry}, Kareem and {Shen}, Rongfeng and {Tam}, Pak-Hin Thomas and {Hu}, Zhecheng and {Yang}, Yanlv and {Zou}, Yuan-Chuan and {Wu}, Jia-Li and {Lei}, Wei-Hua and {Wei}, Jun-Jie and {Wu}, Xue-Feng and {Sun}, Tian-Rui and {Wang}, Fa-Yin and {Zhang}, Bin-Bin and {Xu}, Dong and {Yang}, Yuan-Pei and {Li}, Wen-Xiong and {Xiang}, Dan-Feng and {Wang}, Xiaofeng and {Wang}, Tinggui and {Zhang}, Bing and {Jia}, Peng and {Yuan}, Haibo and {Zhang}, Jinghua and {Xuesong Wang}, Sharon and {Gan}, Tianjun and {Wang}, Wei and {Zhao}, Yinan and {Liu}, Yujuan and {Wei}, Chuanxin and {Kang}, Yanwu and {Yang}, Baoyu and {Qi}, Chao and {Liu}, Xiaohua and {Zhang}, Quan and {Zhu}, Yuji and {Zhou}, Dan and {Zhang}, Congcong and {Yu}, Yong and {Zhang}, Yongshuai and {Li}, Yan and {Tang}, Zhenghong and {Wang}, Chaoyan and {Wang}, Fengtao and {Li}, Wei and {Cheng}, Pengfei and {Shen}, Chao and {Li}, Baopeng and {Pan}, Yue and {Yang}, Sen and {Gao}, Wei and {Song}, Zongxi and {Wang}, Jian and {Zhang}, Hongfei and {Chen}, Cheng and {Wang}, Hui and {Zhang}, Jun and {Wang}, Zhiyue and {Zeng}, Feng and {Zheng}, Zhenhao and {Zhu}, Jie and {Guo}, Yingfan and {Zhang}, Yihao and {Li}, Yudong and {Wen}, Lin and {Feng}, Jie and {Chen}, Wen and {Chen}, Kun and {Han}, Xingbo and {Yang}, Yingquan and {Wang}, Haoyu and {Duan}, Xuliang and {Huang}, Jiangjiang and {Liang}, Hong and {Bi}, Shaolan and {Gai}, Ning and {Ge}, Zhishuai and {Guo}, Zhao and {Huang}, Yang and {Li}, Gang and {Li}, Haining and {Li}, Tanda and {Yuxi} and {Lu} and {Rix}, Hans-Walter and {Shi}, Jianrong and {Song}, Fen and {Tang}, Yanke and {Ting}, Yuan-Sen and {Wu}, Tao and {Wu}, Yaqian and {Yang}, Taozhi and {Yin}, Qing-Zhu and {Gould}, Andrew and {Lee}, Chung-Uk and {Dong}, Subo and {Yee}, Jennifer C. and {Shvartzvald}, Yossi and {Yang}, Hongjing and {Kuang}, Renkun and {Zhang}, Jiyuan and {Liao}, Shilong and {Qi}, Zhaoxiang and {Yang}, Jun and {Zhang}, Ruisheng and {Jiang}, Chen and {Ou}, Jian-Wen and {Li}, Yaguang and {Beck}, Paul and {Bedding}, Timothy R. and {Campante}, Tiago L. and {Chaplin}, William J. and {Christensen-Dalsgaard}, J{\o}rgen and {Garc{\'\i}a}, Rafael A. and {Gaulme}, Patrick and {Gizon}, Laurent and {Hekker}, Saskia and {Huber}, Daniel and {Khanna}, Shourya and {Li}, Yan and {Mathur}, Savita and {Miglio}, Andrea and {Mosser}, Beno{\^\i}t and {Ong}, J.~M. Joel and {Santos}, {\^A}ngela R.~G. and {Stello}, Dennis and {Bowman}, Dominic M. and {Lares-Martiz}, Mariel and {Murphy}, Simon and {Niu}, Jia-Shu and {Ma}, Xiao-Yu and {Moln{\'a}r}, L{\'a}szl{\'o} and {Fu}, Jian-Ning and {De Cat}, Peter and {Su}, Jie and {consortium}, the ET},
        title = "{ET White Paper: To Find the First Earth 2.0}",
      journal = {arXiv e-prints},
         year = 2022,
        month = jun,
          eid = {arXiv:2206.06693},
        pages = {arXiv:2206.06693},
          doi = {10.48550/arXiv.2206.06693},
archivePrefix = {arXiv},
       eprint = {2206.06693},
 primaryClass = {astro-ph.IM},
       adsurl = {https://ui.adsabs.harvard.edu/abs/2022arXiv220606693G}
}

\begin{appendix}

\section{MESA inlists}\label{sec_inlist}

\subsection{Inlist for signal star with mass loss or accretion: inlist\_astero\_gyre }

\begin{verbatim}
&star_job  

change_initial_net = .true. 
new_net_name = 'cno_extras_o18_and_ne22.net' 
create_pre_main_sequence_model = .true.
set_initial_cumulative_energy_error = .true.
new_cumulative_energy_error = 0d0

/ ! end of star_job namelist

&controls

initial_mass = 2.d0 !2.002713d0 !1.d0 
initial_z = 0.02d0
mass_change = -6d-5
max_star_mass_for_gain =1.d0      
!min_star_mass_for_loss = 2d0
accrete_same_as_surface = .true.!      
mixing_length_alpha = 2d0      
max_years_for_timestep = 3d6 !8d6 !3d6
mesh_delta_coeff = 0.5   !0.3 ==> 5100
varcontrol_target = 1d-4
mesh_logX_species(1) = 'h1'      
mesh_logX_min_for_extra(1) = -4      
mesh_dlogX_dlogP_extra(1) = 0.2      
mesh_dlogX_dlogP_full_on(1) = 1d-3     
mesh_dlogX_dlogP_full_off(1) = 1d-5          
calculate_Brunt_N2 = .true.      
smooth_convective_bdy = .true.      
num_cells_for_smooth_brunt_B = 10            
num_cells_for_smooth_gradL_composition_term = 10  
set_min_D_mix = .true.         
mass_lower_limit_for_min_D_mix = 0d0          
mass_upper_limit_for_min_D_mix = 1d99        
min_D_mix = 5d-1 !1d1

overshoot_scheme = 'exponential' ! exponential, step      
overshoot_zone_type = 'any' ! burn_H,_He,_Z, nonburn, any
overshoot_zone_loc = 'core'  !   core, shell, any
overshoot_bdy_loc = 'top'   ! bottom, top, any`
         
overshoot_f = 0.016d0      
overshoot_f0 = 0.001d0
write_pulse_data_with_profile =.TRUE.! .false. !
pulse_data_format = 'GYRE'

/ ! end of controls namelist

\end{verbatim}

\end{appendix}

\end{document}